\documentclass[aoas]{imsart}

\usepackage[utf8]{inputenc}
\usepackage{amsmath}
\usepackage{amsthm}
\usepackage{graphicx}
\usepackage{amsfonts}
\usepackage{subcaption}
\usepackage{float}
\usepackage{bm}
\usepackage{comment}
\usepackage{multirow}
\usepackage{outlines}
\usepackage[ruled, vlined]{algorithm2e}
\usepackage{hyperref}
\hypersetup{
    colorlinks=true,
    linkcolor=blue,
    filecolor=magenta,      
    urlcolor=cyan,
    citecolor=blue
}
\usepackage{bbm}
\usepackage{enumitem}
\usepackage[authoryear, semicolon]{natbib}

\def\Ell{\mathcal L}

\newcommand{\xx}{\mathbf{x}}

\newcommand{\XX}{\mathbf{X}}

\newcommand{\ipsum}{\sum_{i = 1}^p}

\newcommand{\jpsum}{\sum_{j = 1}^p}

\DeclareMathOperator*{\argmin}{arg\,min}
\theoremstyle{plain}
\newtheorem{thm}{Theorem}

\newtheorem{lem}{Lemma}
\newtheorem{cor}{Corollary}

\usepackage[authoryear, semicolon]{natbib}
\begin{document}

\begin{frontmatter}
\title{Subbotin Graphical Models for Extreme Value Dependencies with Applications to Functional Neuronal Connectivity}
\runtitle{Subbotin Graphical Models}

\begin{aug}
\author[A]{\fnms{Andersen} \snm{Chang}\ead[label=e1]{atc7@rice.edu}}
\and
\author[B]{\fnms{Genevera I.} \snm{Allen}\ead[label=e2]{gallen@rice.edu}}
\address[A]{Department of Statistics,  Rice University,
\printead{e1}}

\address[B]{Department of Electrical and Computer Engineering, Rice University,\\
Department of Computer Science, Rice University,\\
Department of Statistics, Rice University,\\
Department of Pediatrics-Neurology, Baylor College of Medicine,\\
Jan and Dan Duncan Neurological Research Institute, Texas Children’s Hospital,
\printead{e2}}
\end{aug}

\begin{abstract}

With modern calcium imaging technology, activities of thousands of neurons can be recorded in vivo. These experiments can potentially provide new insights into intrinsic functional neuronal connectivity, defined as contemporaneous correlations between neuronal activities. As a common tool for estimating conditional dependencies in high-dimensional settings, graphical models are a natural choice for estimating functional connectivity networks. However, raw neuronal activity data presents a unique challenge: the relevant information in the data lies in rare extreme value observations that indicate neuronal firing, rather than in the observations near the mean. Existing graphical modeling techniques for extreme values rely on binning or thresholding observations, which may not be appropriate for calcium imaging data. In this paper, we develop a novel class of graphical models, called the Subbotin graphical model, which finds sparse conditional dependency structures with respect to the extreme value observations without requiring data pre-processing. We first derive the form of the Subbotin graphical model and show the conditions under which it is normalizable. We then study the empirical performance of the Subbotin graphical model and compare it to existing extreme value graphical modeling techniques and functional connectivity models from neuroscience through several simulation studies as well as a real-world calcium imaging data example.

\end{abstract}


\end{frontmatter}

\section{Introduction} \label{sec:intro}

Undirected graphical models are a commonly used unsupervised learning tool for exploring and analyzing network structures and conditional relationships in multivariate data, especially in high-dimensional settings. These models have been used in a wide variety of applications, ranging from genomics \citep{ga1} and network biology \citep{tw1} to finance and manufacturing \citep{mt1}, in order to answer various different research problems. Currently, there exists a broad literature on graphical models on the theoretical properties and estimation procedures for the Gaussian as well as other exponential family distribution graphical models \citep{sl1, jf2, ey3} under the assumption that the observations in the data set are of equal relevance when estimating a graph to represent the dependencies between the features in the data. For some research questions, however, this will not be the case. One particular example where only a portion of observations are of scientific interest is in the problem of extreme value modeling, where the main goal is to find significant relationships between variables with respect to rare extreme value observations rather than the observations close to the mean. Common applications where this is relevant include risk management and variance minimization in finance \citep{ev1} and extreme weather event prediction in climatology \citep{ev2}. In this case, we want the observations corresponding to the extreme values to be given the most influence in the model, while observations close to the mean should be given very little weight if any at all. In particular, for the context of graphical models, our goal for extreme value modeling is to find sparse conditional dependency structures in high dimensional data between features with respect to the occurrences and magnitudes of rare extreme value observations.

\begin{figure}[t]
    \centering
    \begin{subfigure}[t]{0.475\linewidth}
        \includegraphics[width=\linewidth]{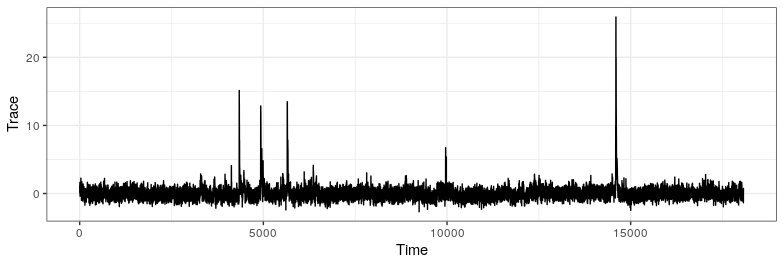}
        \includegraphics[width=\linewidth]{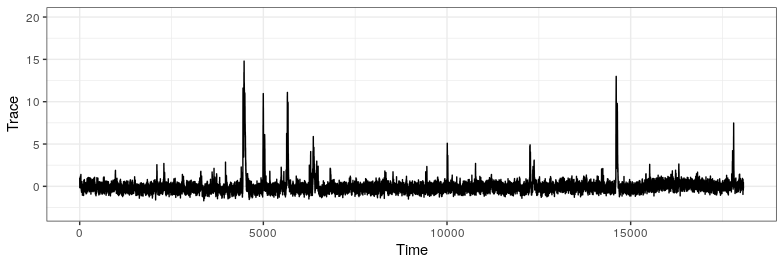}
    \caption{}
    \label{fig:f1}
    \end{subfigure} %
    \begin{subfigure}[t]{0.475\linewidth}
        \includegraphics[width=\linewidth]{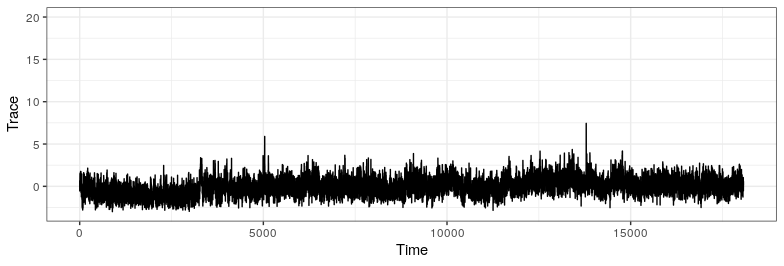}
        \includegraphics[width=\linewidth]{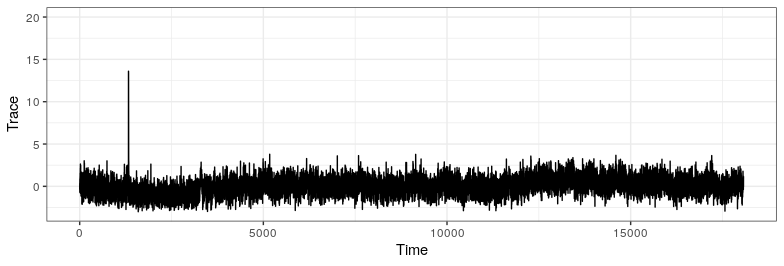}
    \caption{}
        \label{fig:f2}
    \end{subfigure}
    \caption{\textbf{(a)} An example of two fluorescence traces which have contemporaneous extreme value observations. \textbf{(b)} An example of two fluorescence traces which do not have contemporaneous extreme value observations, but are selected as functionally connected when applying a Gaussian graphical model.}
    \label{fig:fl}
\end{figure}

One specific field of interest for which extreme value modeling can be applied is that of neuroscience. In particular, these types of models can be used to help analyze data collected on individual neuronal activity through two-photon calcium imaging \citep{dy1}, which can record simultaneous in vivo recordings of thousands of neurons at a time. For each neuron, a fluorescence trace is recovered that represents neuronal activity over time; we show two examples of these in Figures \ref{fig:f1} and \ref{fig:f2}. The most critical information from the fluorescence trace lies in the times when each neuron is most active in terms of firing events, which are indicated in the data by large positive spikes in the time series. On the other hand, the non-extreme value observations, which are mainly comprised of random noise that occurs in the trace during periods of inactivity, are much more numerous but also not indicative of any useful signal in the data. Because of this particular aspect of the data, any models applied to analyzing calcium imaging data must be focused on fitting to the extreme values that comprise the neuronal spiking events in the data rather than the non-extreme observations in the fluorescence traces; this makes models that are designed fit to the extreme values are a natural fit for analyzing calcium imaging data.

In the context of neuroscience and calcium imaging, graphical models have been applied as a tool to help researchers analyze the connections between individual neuronal activities in the brain \citep{rd1}. These interactions are often viewed in the framework of functional neuronal connectivity, i.e. the statistical relationships between the activity of neurons in the brain \citep{bb1}. Specifically, in this paper, we are interested in applying graphical models in order to find sparse conditional dependency graphs to represent intrinsic functional connectivity, defined as correlated contemporaneous fluctuations in neuronal activity \citep{rb1}, as it can potentially help provide a better understanding how neuronal circuits are organized or be used as a tool for deriving the structural connectivity between individual neurons in the brain. We note that we do not consider directional or time-lagged functional connections in this paper, as calcium imaging data is not sampled at a fast enough rate in order to reliably capture directed causality or time-lagged correlations between neuronal activities, as there can be multiple firing events in each observation in the raw data. In order to obtain meaningful intrinsic functional connectivity estimates from calcium imaging data, the graphical model technique applied should find conditional dependencies between neurons with respect to the occurrences and magnitudes of the rare extreme values in the fluorescence traces, such as in Figure \ref{fig:f1}, as these indicate actual correlations between the neuronal firing activity. On the other hand, correlations with respect to the observations near the mean in fluorescence trace data are typically indicators of measurement artefacts or random noise, rather than actual related firing activity between pairs of neurons. When standard classes of probabilistic graphical models such as exponential family graphical models are applied to raw calcium imaging data, though, the estimated functional connections are often based on correlations with respect to the non-extreme observations; for example, a Gaussian graphical model applied to raw fluorescence traces will consider the pair of neurons in Figure \ref{fig:f2} to be functionally connected, even though there does not appear to be any correlated firing activity between the two. 

Currently, when probabilistic graphical models are applied to calcium imaging data, preprocessing methods are typically used in order to transform the data in order to remove the influence of non-extreme values in model fitting. Using this type of procedure, the raw fluorescence traces collected from a two-photon calcium imaging experiment are denoised to separate the positive extreme value spikes that indicate neuron firing activity from the smaller random fluctuations produced by measurement noise \citep{ep1, jf1}. These methods commonly include applying an automatic thresholding procedure to zero out non-extreme value observations, theoretically leaving only the spikes in the fluorescence traces that would arise from neuronal activity as non-zero observations. Gaussian graphical model methods are then applied to the denoised trace data in order to derive functional neuronal connectivity estimates \citep{dy2, ac1}. While this type of approach is commonly used to analyze calcium imaging data, it suffers from potential drawbacks. By setting all values which considered by the procedure to non-extreme to zero, thresholding inherently binarizes the data between extreme values and non-extreme values. While this can be useful in contexts where a natural threshold exists, the delineation between extreme and non-extreme values in calcium imaging data is not necessarily clear, and any selected threshold may not be particularly interpretable. Additionally, the appropriate threshold in a calcium imaging data set can vary greatly between different individual neurons due to external factors, such as the location of the neuron in the brain relative to the recording device. Thus, algorithmic denoising procedures can zero out an extreme value observation which is scientifically relevant, or leave observations in the data which are actually noise. Adding to this, because the data analysis procedure described here is a multi-step pipeline, errors that arise from the data preprocessing step will propagate to the subsequent graph estimates, thus adding extra unaccounted variability or bias in to the model estimates. Thus, forcing all observations below a threshold to zero and giving it no weight in model estimation may not be the best approach for calcium imaging data. Instead, an approach with increases the influence of extreme values smoothly with respect to the magnitude of the observation may make more sense.

A variety of techniques specific to modeling functional connectivity networks for calcium imaging data have also been developed and applied in the literature as well. These models are not explicitly designed to estimate conditional dependencies with respect to extreme values, but rather assume a distribution on the spiking pattern over time. Several of these models treat the firing times of the neurons as a point process, which can then be modeled as Hawkes model with correlated and autocorrelated conditional intensities \citep{rl1, prb1, me2}. The inter-spike interval times can also be modeled as correlated processes with varying intensities based on linear relationships \citep{mm1}. While these types of models are useful for a variety of contexts, they have several disadvantages for estimating functional connectivity from calcium imaging data. Unlike graphical models, many of these methods do not look at conditional relationships, meaning that they only find bivariate dependencies without simultaneously accounting for the other observed neuronal activities; this can lead to the finding of spurious correlations in the data. These methods are also not designed to perform parameter estimation and model selection at the same time. Instead, a hard threshold is generally applied to the estimated parameters post hoc in order to select edges in the network. This leads to results which are more susceptible to overfitting and less generalizable compared to methods such as Lasso-based estimators \citep{rt1}, does not allow for automatic selection of the number of edges in the final graphical model estimate, and relies on the user to select a threshold after viewing the results rather than through a fully algorithmic, data-driven process. Another issue specific to methods that are based on point processes is that they require the inference of the occurrences of spikes from the raw fluorescence trace data, potentially leading to error propagation in the same manner that can occur when thresholding.

Other types of commonly used methods also utilize ideas that are based on time series modeling. For example, some models assume that the neuronal activity measurements follow a vector autoregressive (VAR) process. VAR models can be applied directly to raw data in order to estimate relationships which accounting for autocorrelation effects simultaneously with model parameters \citep{mk1, ff1}, or a deconvolution algorithm can be applied in order to remove any autoregressive effects from the data \citep{ep1, jv1, sj1} before applying other network models. Models which transform the data to the frequency domain and find correlations in the spectral decompositions of the neuronal activity have been used as well for finding functional connectivity estimates \citep{et1, xg1}. A variety of models which use structural equations in order to replicate neuronal communication have also been proposed \citep{am2}. The assumptions of these models, however, may not be suitable specifically for the problem of estimating intrinsic functional connectivity as we have previously defined from calcium imaging data. The spectral models tend to be tailored towards regular or repeated cyclical fluctuations in the observations, whereas the individual neuronal firing activity as represented by fluorescence traces can potentially occur infrequently and at irregular intervals; thus, the spectral decomposition may not be a good estimator of functional connectivity. The VAR and structural equation models are generally used in order to estimate functional connectivity involving lagged dependencies and possibly directed connections, which we are not considering as part of estimating intrinsic functional connectivity from calcium imaging data in this paper.

Other models that have been introduced include shared information metrics from information theory such as mutual information or joint entropy, which can find nonlinear correlations between neuronal activity which can then be used as directed or undirected functional connectivity estimates \citep{mg1, ks1}, as well as the linear-nonlinear model \citep{sk1, ip1}, which assumes that the spiking patterns for each neuron follow either a Gaussian or Poisson distribution with varying intensities, with the relationship between correlated neurons modeled via a quadratic function. Both of these types models have shown to be useful for a variety of research problems in neuroscience, such as estimating the relationship between neuronal activity and stimuli or estimating correlated neuronal activity in smaller data sets. However, they have several potential disadvantages for estimating a full conditional dependency structure to represent functional connectivity in high dimensions. Like the point process models discussed previously, these methods do not allow for the simultaneous parameter estimation and model selection through a data-driven approach, instead relying on post-hoc procedures to select the functional connections. Additionally, these methods can potentially be computationally slow for high dimensional data; specifically, the linear-nonlinear model can be difficult to estimate if there is no closed-form solution for the joint likelihood function, and the mutual information models require estimating empirical joint and conditional distributions for each pair of features in the data.

Our goal for this paper is to design a model which allows us to estimate functional connectivity networks from calcium imaging data under the graphical modeling paradigm, without the disadvantages of the existing methods. We introduce a new class of probabilistic graphical model which can be used to find conditional dependencies between variables based on extreme value observations. The basis of the graphical model is the Subbotin distribution, also known as the generalized normal distribution \citep{sn1}. With any power greater than 2, the Subbotin distribution will have thinner tails when compared to a Gaussian distribution, which in turn comparatively increases the weight of large differences between an observation and its estimated expected value in the corresponding likelihood function. This means that parameter estimates from a Subbotin distribution with power greater than 2 will be more heavily weighted towards the observed extreme values of a data set, relative to parameter estimates from a Gaussian distribution. Thus, we propose to use a construction of a multivariate Subbotin distribution with a power greater than 2 as a graphical model distribution for analyzing data sets where the goal is to find relationships between variables based on their extreme value observations. Compared to models currently used in the literature for graphical modeling for extreme values, the Subbotin graphical model has several advantages; it does not require the usage of any data pre-processing, and it smoothly increases the relative weight of extreme value observations in model fitting by their magnitude instead of binarizing data as extreme or non-extreme. It also allows for simultaneous parameter estimation and sparse model selection via regularization, which many current functional connectivity network estimation methods do not include. Through empirical data studies, we compare our proposed method to other extreme value graphical model methods, and we show that the Subbotin graphical model generally performs better than existing models on the graph edge recovery problem for extreme value observations in high dimensional data.

The rest of this paper is structured as follows. In section 2, we introduce the Subbotin graphical model by defining its node-wise conditional distribution, and derive the joint multivariate distribution of the graphical model using results from the Hammersley-Clifford theorem. In section 3, we introduce the estimation algorithm and explore its theoretical properties. In section 4, we investigate the performance of the Subbotin graphical model on simulation studies and compare it to state-of-the-art methods. Finally, in section 5, we apply our method to a real-world calcium imaging data set to estimate a functional neuronal connectivity graph and explore the results.

\section{Subbotin Graphical Models} \label{sec:meth}

\subsection{Background: Extreme Value Graphical Models} \label{sub:evgm}

Suppose we have a random data matrix $\XX \in \mathbb{R}^{n \times p}$ with constant mean and variance for which we would like to estimate pairwise conditional relationships between the features, specifically with respect to the observed extreme values. In this particular context, we presume that the extreme value events are empirically rare relative to the number of total observations, and therefore in order to see extreme events in the data we need the number of observations $n$ to be large relative to the number of variables $p$. Several approaches which can be used to find conditional dependencies between observed extreme values, but which have not been applied to calcium imaging data, have been proposed in graphical modeling literature. Many of these utilize a multivariate extension of the generalized extreme value (GEV) distribution, which has been shown to be the limiting distribution of the maxima of normalized i.i.d. variables \citep{tb1}. From this, several different methods for extreme value modeling have been developed. One of these is the block maxima approach, which assumes the maxima within predetermined time bins of each variable all follow a marginal univariate GEV distribution. The parameters of the GEV distributions corresponding to each variable are estimated separately using maximum likelihood estimation on the block maximal data. These are then copularized based on the estimated probability density function and converted to Gaussian distributed data, after which a Gaussian graphical model is used to estimate the conditional dependency structure  \citep{hy1}. However, one particular issue with the block maxima model for this problem is that it relies on the choice of the size of each time bin. While the method may lend itself well to data with natural cutoff points, e.g. the maximum daily rainfall in various locations in each year. However, neuronal activity does not have this kind of property, as there may be long periods of time in which a neuron is inactive alongside short periods of large activity. Thus, converting fluorescence trace data from the raw measurements to block maxima could remove relevant information in the form of smaller extreme values in the same time bin, or there could be many block maxima which correspond to noise from the fluorescence trace instead of actual neuronal activity.

Another family of models utilize a peaks-over-threshold method, in which it is assumed that the distribution of all values above a predetermined threshold follow a GEV distribution. Here, only the observations in which there was at least one feature with an extreme value are kept; a multivariate Pareto distribution is then fit to the remaining data. It has been shown that the parameterization of certain classes of the multivariate Pareto distribution can be used to represent the conditional dependency structure between features with respect to the extreme values in the case tree-structured graphs and for the H\"{u}sler-Reiss multivariate extreme value distribution \citep{se1}. Graph estimation for the model can be performed in one of several ways: parameter estimates can be found for a set of a priori selected edges in the underlying graph using the likelihood of the distribution, or a step-wise procedure with a penalized likelihood metric can be used in order to select edges and estimate the parameters. While this approach has been shown to be useful in various applications, such as climatology and environmental statistics, calcium imaging data presents unique challenges that create difficulties for this type of method. The peaks-over-threshold model requires a choice on the threshold for which a value is considered extreme, which leads to the same issues as the data pre-processing when fitting a Gaussian or exponential family graphical model discussed in Section \ref{sec:intro}. The potential rarity of the relevant extreme values that can occur in a single fluorescence trace can lead to issues with small sample sizes when estimating the parameters of the multivariate Pareto distribution, as the procedure removes all observations below the threshold selected a priori. Additionally, the estimation procedure for the formulation for the multivariate Pareto model is restricted by technical limitations. An a priori graph structure for a functional neuronal connectivity network is not generally known, meaning that selecting particular edges for the graph before estimation does not make sense for this problem. The estimation algorithm is also only of practical use for low-dimensional data, graphs with small cliques, or for tree structured graphs \citep{se3}. Thus, this model is difficult to use in modern calcium imaging data as there can be hundreds or thousands of neurons measured simultaneously and because functional connectivity networks are generally not restricted to block or tree-structured graphs.

\begin{figure}[t] 
    \centering
    \medskip
    \includegraphics[scale = 0.6]{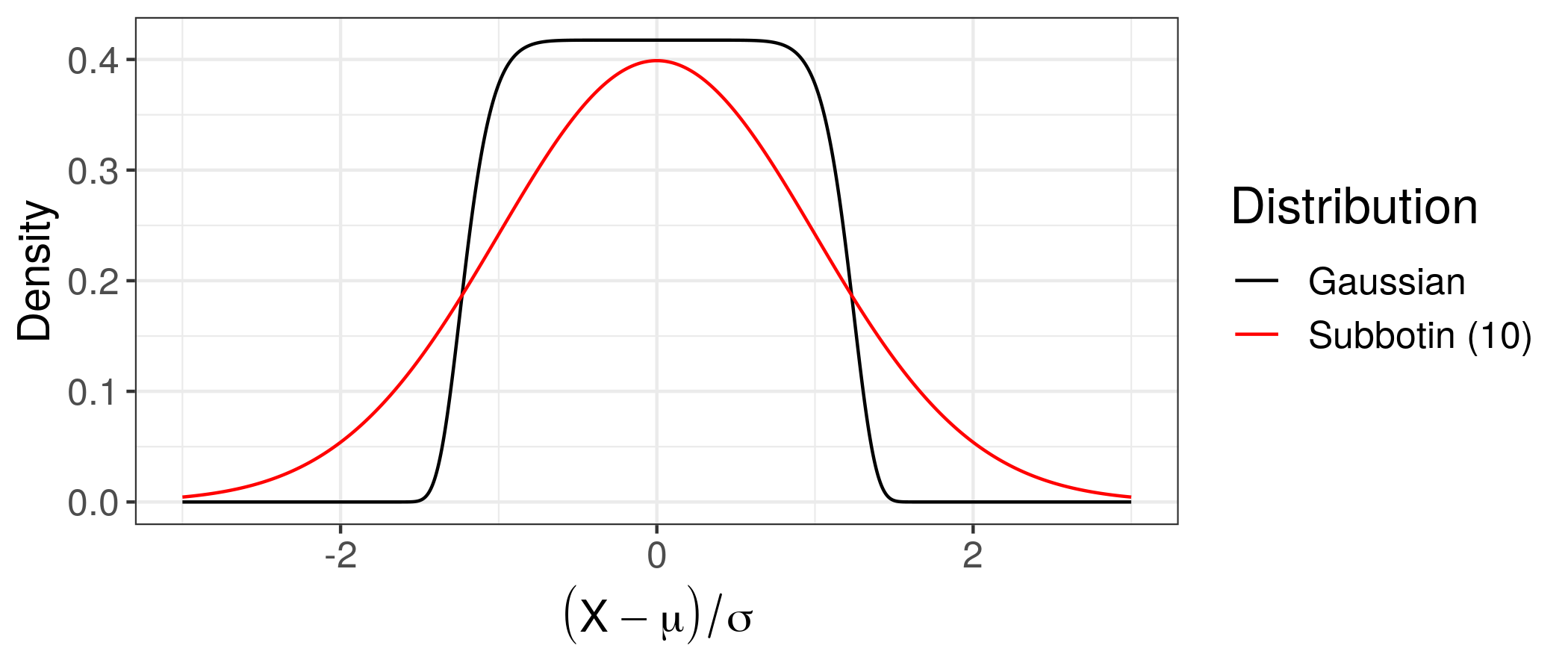}
    \caption{Density functions for the Gaussian distribution (red) and Subbotin distribution with $\nu = 10$ (black).}
    \label{fig:sub}
\end{figure}

A potential alternative solution for finding a graphical model for extreme values which does not utilize the GEV distributional assumption is the quantile graphical model \citep{aa1}, specifically used at a very high or very low quantile in order to fit to the observed extreme values. The quantile graphical model finds associations between variables at a specified quantile of the data using a weighted $\ell_1$ loss function. This potentially removes the need to pre-process the data, as the quantile graphical model will increase the relative influence of extreme values in a data set compared to a Gaussian graphical model if an appropriate quantile is chosen. However, the selection of the appropriate quantile to use for a particular data set can be extremely difficult, as it is not perfectly clear how to select a specific quantile to use when the number and magnitude of the extreme values can vary substantially between the each of the variables in a data set.

\subsection{Subbotin Graphical Model} \label{sub:sgm}

Because of the aforementioned disadvantages of the current extreme value graphical model methods, our goal is to define a new class of graphical model that can find sparse conditional dependency structures in high dimensional data with respect to extreme value observations, but which does not require either thresholding or binning the data or a definition of an extreme value a priori. To create a model that can find relationships between extreme values, we draw inspiration from the opposite statistical problem, namely that of robust modeling, in which parameter estimates that are insensitive to the presence of outliers in the data are desired. The basis for many robust models is the Subbotin, or generalized normal, distribution, with location parameter $\mu$, scale parameter $\sigma$, and shape parameter $\nu$ of the form \begin{equation} \label{eq:usub}
    f(x) = \frac{1}{2 \sigma \Gamma\left(\frac{1+\nu}{\nu}\right)}e^{-\left(\frac{|x - \mu|}{\sigma}\right)^{\nu}}, \, x \in \mathbb{R}.
\end{equation} If $\nu = 2$, then Subbotin distribution is proportionally equivalent to the Gaussian distribution. Robust models use the Subbotin distribution with $0 < \nu < 2$; this has been shown to produce results that are more stable in the presence of potential extreme values compared to Gaussian-based models \citep{he1, rc1}. Thus, in order to create a model that more heavily weights extreme values, we propose to use the opposite class of distribution, i.e. the Subbotin distribution with $\nu > 2$. We show an example of this distribution in Figure \ref{fig:sub}, comparing the Subbotin with $\nu = 10$ to a Gaussian distribution. This concept has been applied in the case of linear regression modeling; for example, $\ell_{p}$ norm Lasso regression with $p > 2$ has been applied a method for feature selection with respect for extreme values \citep{ac2}, and $\ell_{\infty}$ norm regression has been used to minimize the maximum error in prediction \citep{hn1, am1}. The Subbotin distribution with $\nu > 2$ will be more sensitive to values that are far from the mean compared to a Gaussian, as for any $\nu > 2$, the Subbotin distribution will have thinner tails relative to the Gaussian; thus, any large magnitude values of $x$ will have a lower probability density relative to the Gaussian. Therefore, in this setting, any extreme value observations of $x$ will have a much larger impact on any parameter estimates, since the Subbotin model with $\nu > 2$ assumes that these extreme values are much less likely to occur when compared to the Gaussian case. We note that the Subbotin distribution is not a generative distribution for extreme values, as it is a thin-tailed distribution with very low density outside of one $\sigma$ away from the mean; in particular, a larger $\nu$ will actually lead to a generative distribution with fewer extreme values. Instead, we choose this distribution because parameter estimates of $\mu$ and $\sigma$ will be more heavily influenced by extreme values when compared to distributions such as the Gaussian, especially with larger values of $\nu$.

In order to apply the Subbotin distribution as a graphical model, we first construct a multivariate Subbotin distribution based on the univariate distribution. Different definitions of a multivariate Subbotin distributions have been proposed and explored in previous works. \citep{sds1} first introduced a multivariate Subbotin distribution, studied later in several other works \citep{eg1, gv1}, of the form 
\begin{equation}
    f(\mathbf{y}) = \frac{p\Gamma(p/2)}{\pi^{p/2} \Gamma(1 + p/2\beta) 2^{1 + p/2\beta}} \exp\left( -\frac{1}{2}[(\mathbf{y}-\mu)^T \Sigma^{-1} (\mathbf{y} - \mu)]^{\beta}\right).
\end{equation} This distribution essentially raises the exponent of a multivariate Gaussian distribution to a power of $\beta / 2$, with a new normalizing constant for arbitrary $\beta$. \citep{ig1} later introduced a more general multivariate generalized normal distribution, of the form \begin{equation}
    f(\mathbf{y}) = \frac{1}{(2\Gamma(1 + 1 / \theta )) ^ p \det(\mathbf{C})}\exp\left( - \ipsum |\jpsum \mathbf{C}_{ij} (y_j - \mu_j) |^{\theta}\right)
\end{equation} where $\mathbf{C}$ is a possibly non-symmetric parameter matrix. While both of these distributions can be useful in different contexts and models, these formulations do not work well as definitions for a graphical model distribution. In particular, these multivariate distributions will contain multi-way interactions between features which can change in complexity based on the shape parameter selected for the distribution, whereas we want to limit ourselves exclusively to pairwise conditional dependency interpretations in our graphical model. Thus, we require a new definition of the multivariate Subbotin distribution. 

To define this distribution, we follow the methodology utilized in \citep{jb1} and \citep{ey1}. In these papers, the form of the graphical model distribution is established by first defining the node-wise conditional distributions, and then using them to construct the joint multivariate distribution. For the Subbotin graphical model, we propose to use the Subbotin distribution as the node-wise conditional distribution for each individual feature. Because of the aforementioned properties, we choose to use the Subbotin with $\nu > 2$ as the node-wise conditional distribution for the Subbotin graphical model. Throughout the rest of this paper, we let $\mu = 0$ and $\sigma = 1$ without loss of generality. We will also restrict ourselves to the case where $\nu$ is an even integer. With these particular parameter values set, we can cast Equation \ref{eq:usub} as a node-wise conditional distribution of the form
\begin{equation} \label{eq:csub}
    f(x_i|\xx_{-i}) = \frac{1}{2 \Gamma\left(\frac{1+\nu}{\nu}\right)}e^{-\left( \theta_{ii} x_i - (\sum_{j \neq i} \theta_{ij}x_j)\right)^{\nu}}, \, x_i \in \mathbb{R}, \, \nu > 2.
\end{equation} Here, the conditional relationships between a pair of variables $x_i$ and $x_j$ are assumed to be linear in nature and are represented by the parameter $\theta_{ij}$. If the parameter $\theta_{ij}$ is zero, then it is trivial to show that $x_i$ and $x_j$ are conditionally linearly independent given all of the other variables in $\XX$. Just as for Equation \ref{eq:usub}, the estimates of the model parameters $\theta_{ij}$ will be more heavily weighted towards fitting to the extreme value observations of $x_i$ when $\nu > 2$ relative to a Gaussian distribution. In particular, we can treat $\nu$ as a tuning parameter that controls how much influence the extreme value observations will have on the model estimates.

For the cases which are applicable for the Subbotin graphical model distribution, i.e. when $\nu > 2$, the resulting node-wise conditional distribution is a special case of a curved exponential family distribution, as opposed to the more commonly used exponential family distribution. Because of this, our method is not conducive to straightforward applications of results from the exponential family graphical model literature. Also, as a whole, properties of probabilistic graphical model distributions stemming from this type of node-wise conditional distribution have not been well-studied in the literature. Therefore, in order to formulate the joint graphical model distribution for the Subbotin graphical model, we utilize the most general method as originally postulated in \citep{jb1}. In that paper, the author shows how joint multivariate graphical model distributions can be constructed from any set of valid node-wise conditional distributions by applying the most general form of a valid graphical model distribution, as proven via the Hammersley-Clifford theorem \citep{jh1}. Here, we apply these same ideas specifically for the Subbotin node-wise conditional distribution in Equation \ref{eq:csub} in order to derive the joint graphical model distribution. The result is shown in the following theorem.

\begin{thm}[Joint Distribution of the Multivariate Subbotin]
\label{thm:jd}
Suppose ($x_1, x_2, \hdots , x_p$) is a p-dimensional random vector, and its node-wise conditional distributions are given by: $$f(x_i|\xx_{-i}) = \frac{1}{2\Gamma(\frac{1+\nu}{\nu})} e^{-(\theta_{ii} x_i - \sum_{j \neq i}\theta_{ij} x_{j})^{\nu}}, \, x_i \in \mathbb{R}$$ for some even integer $\nu$. Then the joint graphical model distribution is given by 
\begin{equation} \label{eq:jsub}
    f(\xx) = \exp \left(\sum_{i = 1}^p \left(- (\theta_{ii} x_i - \sum_{j < i} \theta_{ij}x_j)^{\nu} + (\sum_{j < i} \theta_{ij}x_j)^{\nu} \right) - A(\boldsymbol{\Theta}) \right)
\end{equation} where $\boldsymbol{\Theta}$ is the parameter matrix, i.e. the collection of all model parameters $\theta_{ij}$.
\end{thm} \noindent We note that Theorem \ref{thm:jd} also assumes without loss of generality that the marginal mean of each variable is 0 and the marginal variance is 1; in practice, each feature can be centered and scaled individually using their empirical distributions in order to match these assumptions. In the case where $\nu = 2$, the distribution given by Equation \ref{eq:jsub} is proportional to the multivariate Gaussian distribution, which we formalize below in Corollary \ref{cor:norm}. 

\begin{cor}[Gaussian Equivalence]
\label{cor:norm}
Consider the Subbotin graphical model distribution from Equation \ref{eq:jsub}. Let $\nu = 2$. Then the Subbotin graphical model distribution will be equivalent to the multivariate Gaussian distribution with inverse covariance matrix $\boldsymbol{\Theta}$.
\end{cor} 

In general, for the Subbotin graphical model, the log normalizing constant for the joint distribution $A(\boldsymbol{\Theta})$ does not have a closed form solution. However, using the estimation procedure described in section \ref{sec:mod}, we do not need to have a closed form solution for $A(\boldsymbol{\Theta})$ in order to estimate the graph structure for the joint graphical model distribution, as long as we have a closed form for each of the node-wise conditional distributions. Also, while we do not know the analytical solution for $A(\boldsymbol{\Theta})$, we can still find the condition under which it will be finite and positive and thus when full graphical model distribution will exist; this result is stated in Theorem \ref{thm:norm} below.

\begin{thm}[Normalizing Condition]
\label{thm:norm}
Consider the Subbotin graphical model distribution from Equation \ref{eq:jsub}. Define the matrix $\boldsymbol{\Theta}$ such that $$\boldsymbol{\Theta}_{ij} = \begin{cases} 
     \theta_{ii} & i = j \\
    - \theta_{ij} & i \neq j \\
  \end{cases}.$$ Then the distribution will be normalizable if and only if $\boldsymbol{\Theta}$ is positive definite.
\end{thm}
\noindent We note that the statement of Theorem \ref{thm:norm} shows that the normalizing condition for the Subbotin graphical model distribution is the same as that for the multivariate Gaussian distribution, regardless of which value of $\nu$ is chosen. Full proofs of Theorems \ref{thm:jd} and \ref{thm:norm} can be found in Sections B and C of the Appendix, respectively.

\subsection{Subbotin Graphical Model Selection} \label{sec:mod}

To estimate the edge set of the graph for a given $\nu$, we apply the neighborhood selection approach for sparse graphical model selection \citep{nm1, pr1}, which utilizes generalized linear models in order to estimate the neighbor set of each node. For the Subbotin distribution with power $\nu$, the corresponding generalized linear model is the $\ell_{\nu}$-norm linear regression model \citep{am1}, i.e. $$\hat{\boldsymbol{\theta}}_{ij} = \argmin_{\boldsymbol{\theta}_{ij}} \frac{1}{\nu N}\|\mathbf{x_i} - \mathbf{X_{-i}} \boldsymbol{\theta}_{ij} \|_{\nu}^{\nu}$$ where $\mathbf{X_{-i}}$ denotes all columns of $\XX$ except for column $i$. An $\ell_1$ regularization penalty is added to the regression problem in order to induce sparsity in the selected number of relevant features, giving us as the likelihood function for neighborhood selection $$ \hat{\boldsymbol{\theta}}_{ij} =  \argmin_{\boldsymbol{\theta}_{ij}} \frac{1}{\nu N}\|\mathbf{x_i} - \mathbf{X_{-i}} \boldsymbol{\theta}_{ij} \|_{\nu}^{\nu} + \lambda \|\boldsymbol{\theta}_{ij}\|_1. $$ This is equivalent to the Extreme Lasso regression model as defined by \citep{ac2}, and the parameters for the regression problem can be estimated using the algorithmic procedure outlined in the aforementioned paper. Code which implements the estimation procedure for the Subbotin graphical model has also been available online \citep{ac4}.

Below, we present the finite sample performance guarantees for estimation of the Subbotin graphical model. These results follow directly from the those of \citep{ac2}, in which the authors show that the Extreme Lasso is variable selection consistent with a known convergence rate. Formally, let us define the true edge set of the full graph as $$E = \{(i, j): \, \boldsymbol{\Theta}_{ij} \neq 0\}$$ and the true set of neighbors of each node $i \in \{1, 2, \hdots, p\}$ as $$S_i = \{j \in \{1, 2, \hdots, p\} \setminus i: \,(i, j) \in E\}.$$ We also denote the Fisher information matrix of the joint multivariate distribution as $Q$ and the Fisher information sub-matrix for the true set of neighbors of each node as $Q_{S_iS_i}$. For the theorem below, we require the following conditions:

\begin{enumerate}
    \item \textbf{Bounded eigenvalues}: For all $i \in \{1, 2, 3, \hdots, p\}$, there exists some $B_{min} > 0$ and $B_{max} > 0$ such that $$B_{max} > \Lambda_{\min}(Q_{S_iS_i}) > B_{min}.$$ 
    \item \textbf{Incoherence}: There exists a constant $\alpha >0$ such that $$\|Q_{S_i^c S_i}Q_{S_iS_i}^{-1}\|_{\infty} < 1-\alpha.$$
    \item \textbf{$\theta$-min}: Let $$\min_{(i, j) \in E} |\theta_{ij}| > (\frac{\tau}{\kappa_{\text{IC}}} \cdot \frac{1}{4} + 1 ) \cdot \frac{2\sqrt{s}}{\kappa_{ \Ell}} \cdot \frac{4 \kappa_{\text{IC}}}{\tau} \nu \sqrt{ \frac{\log p}{n}} \bigg[    2 \sqrt{ \frac{2}{\nu}}   +   \sqrt{ \frac{\log p}{n}}   \bigg].$$
\end{enumerate}  With these, we can now state the model selection consistency results of our method.
\begin{thm}[\textbf{Graphical Model Selection Consistency}]  \label{subbotinmodelconsis} Consider the Subbotin graphical model distribution from Equation \ref{eq:jsub}. Assume that Conditions 1 and 2 hold. Suppose the regularization parameters for the estimation procedure satisfy $$\lambda_n \geq \frac{4 \kappa_{\text{IC}}}{\tau} \nu \sqrt{ \frac{\log p}{n}} \bigg[    2 \sqrt{ \frac{2}{\nu}}   +   \sqrt{ \frac{\log p}{n}}   \bigg]$$ where $\kappa_{\text{IC}}$ is the  compatibility constant defined in \citep{lee}. Then, the following properties holds with probability greater than   $1 -  c_1 \exp(- c_2  \log p)$: 

(i) The estimated edge set has a unique solution with support contained
within the true edge set, i.e. $\hat E \subset E$.

(ii) Additionally, if Condition 3 holds, then $\hat{E}$ is also sign consistent, i.e. $\text{sign} (\hat{E}) = \text{sign} (E)$.
\end{thm} \noindent The full proof of Theorem \ref{subbotinmodelconsis} can be found in Section D of the Appendix. We note that the concentration bound for the Subbotin graphical model is a substantially weaker bound compared to that for a Gaussian graphical model with a neighborhood selection approach; this means that the Subbotin graphical model will generally require a relatively larger number of samples in order to achieve the same level of accuracy. However, as mentioned previously, we expect that the number of observations $n$ will be large relative to the number of variables $p$ in the contexts where we would apply the Subbotin graphical model. Thus,  the slower convergence rate should not have a substantial impact empirically on the reliability of the estimates from the Subbotin graphical model. 

As part of the model selection process, values for the hyperparameters $\nu$ and $\lambda$, which control the influence of extreme values and the sparsity level of graph, respectively, need to be chosen as well. A number of methods for graph hyperparameter selection have been studied in the literature; for the Subbotin graphical model, we propose to use a stability selection approach \citep{nm2}. With this method, we utilize a bootstrapping approach to find the probability that each possible edge in the graph will be selected in a graph estimate using random resampling of data for given values of $\nu$ and $\lambda$. Individual edges can then be selected based on a pre-selected stability threshold, or an overall stability metric for each pairing of $\nu$ and $\lambda$ can be calculated using a criterion, such as the one proposed in \citep{hl1}. Final hyperparameter values can then be selected by using the set that gives the largest set of stable edges as the final graph estimate.

\section{Simulation Studies} \label{sec:sim}

We now study the performance of the Subbotin graphical model on several simulations and compare it to various other graphical modeling techniques. For each simulation set, we show the average and standard deviation of the F1 scores for each model with respect to estimating the true edge set of the underlying graph of the data across 50 replications. Below, we consider the Subbotin graphical model for $\nu = 4, 6 $ and $8$, the Gaussian graphical model estimated using neighborhood selection and the graphical Lasso, the quantile graphical model at the 50th and 90th percentiles, and the block maxima copula graphical model at bin sizes of 10, 20, and 30. We note that we do not include the multivariate Pareto graphical model in our simulation comparisons because of the restriction to either tree-structured graphs or to low-dimensional data. We analyze the performance when hyperparameter selection is performed for all methods using oracle sparsity tuning, i.e. we select the number of edges in the estimates from each model to match the true number of edges in the underlying graph from which the data is simulated. Each of the methods are compared using F1 scores with respect to the edges in the true underlying graph used to create the simulated data.

\subsection{Subbotin Graphical Model Distribution} \label{sub:sub}

We first investigate the efficacy of our estimation procedure on the multivariate graphical model distribution from which it is derived, in order to show that our model and estimation procedure are empirically valid. We generate data from the Subbotin graphical model distribution as defined in Equation \ref{eq:jsub}, using a Gibbs sampling procedure with the univariate node-wise conditional distributions as defined in Equation \ref{eq:csub}. The model parameter matrix $\boldsymbol{\Theta}$ is structured as a sparse small world graph with 5 total cliques. By default, we create data sets of 2000 observations of 100 features from a block graph from the Subbotin graphical model distribution with $\nu = 8$; we then vary the number of observations, the number of features, the $\nu$ parameter of the Subbotin graphical model distribution, and the structure of the underlying graph. We also show results for these simulation studies using data-driven stability selection hyperparameter tuning in Section A.1 in the Appendix.

\begin{table}[t]
\caption{Average F1 scores (std. devs) for multivariate Subbotin graphical model distribution simulations with oracle tuning, averaged over 50 replicates. Best performing methods are boldfaced. The best performing model is the Subbotin graphical model with a correctly specified $\nu$, which we would expect since it is the generative distribution for the simulations.}
\label{tab:mvgn}
\centering
\begin{footnotesize}
\begin{tabular}{|l|r|r|r|r|}
    \hline
    \textbf{Model}  & $\boldsymbol{n = 500, p = 100}$ & $\boldsymbol{n = 1000, p = 100}$ & $\boldsymbol{n = 2000, p = 100}$ & $\boldsymbol{n = 4000, p = 100}$ \\
    \hline
    Subbotin (4) & 0.591 (0.062) & 0.642 (0.055)  & 0.787 (0.055) & 0.844 (0.049) \\
    Subbotin (6) & 0.643 (0.066) & 0.671 (0.056) & 0.811 (0.056) & 0.856 (0.052) \\
    Subbotin (8) & \textbf{0.705} (0.067) & \textbf{0.762} (0.059) & \textbf{0.834} (0.055) & \textbf{0.887} (0.050) \\
    \hline
    Gaussian NS & 0.532 (0.060) & 0.549 (0.056) & 0.732 (0.050) & 0.844 (0.050) \\
    Gaussian Glasso & 0.535 (0.057) & 0.551 (0.056) & 0.736 (0.052) & 0.850 (0.049) \\
    \hline
    Quantile (0.5) & 0.478 (0.054) & 0.566 (0.077) & 0.657 (0.064) & 0.822 (0.064) \\
    Quantile (0.9) & 0.211 (0.062) & 0.244 (0.080) & 0.249 (0.070) & 0.351 (0.067) \\
    Quantile (0.99) & 0.012 (0.003) & 0.021 (0.005)  & 0.024 (0.004) & 0.022 (0.003) \\
    \hline
    Copula (10) & 0.512 (0.066) & 0.534 (0.060) & 0.550 (0.056) & 0.535 (0.047) \\
    Copula (20) & 0.497 (0.062) & 0.521 (0.057) & 0.542 (0.056) & 0.526 (0.044) \\
    Copula (30) & 0.465 (0.063) & 0.500 (0.058) & 0.531 (0.052) & 0.520 (0.043) \\
    \hline
    \hline
    \textbf{Model}  & $\boldsymbol{n = 2000, p = 30}$ & $\boldsymbol{n = 2000, p = 100}$ & $\boldsymbol{n = 2000, p = 300}$ & $\boldsymbol{n = 2000, p = 500}$ \\
    \hline
    Subbotin (4) & 0.796 (0.040) & 0.787 (0.055) & 0.660 (0.059) & 0.527 (0.058) \\
    Subbotin (6) & 0.816 (0.042) & 0.811 (0.056) & 0.703 (0.058) & 0.594 (0.060) \\
    Subbotin (8) & \textbf{0.845} (0.039) & \textbf{0.834} (0.055) & \textbf{0.738} (0.060) & \textbf{0.636} (0.055) \\
    \hline
    Gaussian NS & 0.801 (0.030) & 0.752 (0.050) & 0.586 (0.052) & 0.520 (0.051) \\
    Gaussian Glasso & 0.808 (0.031) & 0.766 (0.052) & 0.589 (0.053) & 0.523 (0.051) \\
    \hline
    Quantile (0.5) & 0.722 (0.033) & 0.657 (0.064) & 0.573 (0.078) & 0.443 (0.077) \\
    Quantile (0.9) & 0.243 (0.052) & 0.249 (0.070) & 0.205 (0.081) & 0.205 (0.079) \\
    Quantile (0.99) & 0.023 (0.008) & 0.021 (0.005) & 0.012 (0.004) & 0.010 (0.004) \\
    \hline
    Copula (10) & 0.581 (0.054) & 0.550 (0.056) & 0.524 (0.050) & 0.461 (0.059) \\
    Copula (20) & 0.562 (0.051) & 0.542 (0.056) & 0.525 (0.053) & 0.459 (0.060) \\
    Copula (30) & 0.533 (0.050) & 0.531 (0.052) & 0.521 (0.057) & 0.453 (0.060) \\
    \hline
    \hline
    \textbf{Model}  & $\boldsymbol{\nu = 4}$ & $\boldsymbol{\nu = 6}$ & \textbf{Chain Graph} & \textbf{Erdos-Renyi Graph} \\
    \hline
    Subbotin (4) & \textbf{0.886} (0.054) & 0.821 (0.059) & 0.688 (0.051) & 0.624 (0.060) \\
    Subbotin (6) & 0.832 (0.056) & \textbf{0.863} (0.058) & 0.715 (0.053) & 0.634 (0.064) \\
    Subbotin (8) & 0.733 (0.060) & 0.766 (0.055) & \textbf{0.721} (0.054) & \textbf{0.681} (0.071) \\
    \hline
    Gaussian NS & 0.869 (0.052) & 0.777 (0.049) & 0.653 (0.048) & 0.615 (0.059) \\
    Gaussian Glasso & 0.866 (0.050) & 0.779 (0.050) & 0.657 (0.047) & 0.619 (0.057) \\
    \hline
    Quantile (0.5) & 0.656 (0.069) & 0.611 (0.070) & 0.588 (0.066) & 0.545 (0.069) \\
    Quantile (0.9) & 0.266 (0.075) & 0.244 (0.080) & 0.209 (0.074) & 0.223 (0.078) \\
    Quantile (0.99) & 0.027 (0.003) & 0.021 (0.004)  & 0.010 (0.002) & 0.009 ($<$ 0.001) \\
    \hline
    Copula (10) & 0.566 (0.050) & 0.554 (0.053) & 0.551 (0.051) & 0.460 (0.055)\\
    Copula (20) & 0.557 (0.052) & 0.538 (0.054) & 0.552 (0.052) & 0.477 (0.055)\\
    Copula (30) & 0.552 (0.053) & 0.530 (0.050) & 0.548 (0.053) & 0.455 (0.052)\\
    \hline
\end{tabular}
\end{footnotesize}
\end{table}

Table \ref{tab:mvgn} shows the results from the simulation study. In general, the Subbotin neighborhood selection algorithm is able to identify the true network structure of the Subbotin distribution with fairly high accuracy when the correct parameter value of $\nu$ is specified. We also see that the Gaussian graphical models and the median quantile graphical model performs relatively well in terms of model selection, while the quantile graphical model at the 90th percentile generally performs much worse. These results are unsurprising, as they reflect the fact that the multivariate Subbotin distribution of Equation \ref{eq:jsub} is not itself a generative distribution for extreme values. Specifically, since the probability density function of the Subbotin graphical model distribution is a symmetric bell-shaped curve around the location parameter, we would expect the graphical models which fit toward the mean or the median to perform well on data generated from the distribution. On the other hand, sampling from the multivariate Subbotin distribution is unlikely to generate contemporaneous extreme values between two variables which are related in the parameter matrix, as it is a thin tailed distribution. Thus, the quantile graphical models with high quantiles are unlikely to estimate the true network structure well, which we see reflected in the results. However, the copula block maxima model still performs fairly well even though the data do not come from an extreme value distribution, which likely is because the block maxima from the Subbotin graphical model distribution follow the GEV distribution which underlies the model. One other notable aspect of the results is that, when $\nu$ is misspecified, the Subbotin graphical model sometimes does not perform as well as the Gaussian graphical model; in particular, with larger ratios of the number of observations to the number of features, the Gaussian graphical model has a slightly higher F1 score compared to the Subbotin graphical model, even though the latter has $\nu$ closer to the true $\nu$. This could be because of the sensitivity of the model to outliers, which is reflected in the slower theoretical rate of convergence with respect to observations relative to the Gaussian, as seen in Theorem \ref{subbotinmodelconsis}. Comparing the results for simulation parameters, we see that reducing the number of observations and increasing the number of features decreases the true positive rate for all methods, with a greater impact on the Gaussian graphical model in the former and the Subbotin graphical model in the latter. We also see that the Subbotin graphical model with the true $\nu$ always has the best performance, which we would expect. Finally, we see that the Subbotin graphical model does best for all the various types of graph structures, although the true positive rate decreases more severely for non-small world graphs.

\subsection{Block Maxima Extreme Value Distribution} \label{sub:bm}

\begin{table}[t]
\caption{Average F1 scores (std. dev.) for the block maxima simulations with oracle tuning, averaged over 50 replicates. Best performing methods are boldfaced. The block maxima copula graphical model does best, as it is the generative distribution for the data. The Subbotin graphical model with $\nu = 4$ performs almost as well.}
\label{tab:bm}
\centering
\begin{footnotesize}
\begin{tabular}{|l|r|r|r|r|}
    \hline
    \textbf{Model}  & $\boldsymbol{n = 5000, p = 25}$ & $\boldsymbol{n = 5000, p = 50}$ & $\boldsymbol{n = 5000, p = 100}$ & $\boldsymbol{n = 5000, p = 200}$ \\
    \hline
    Subbotin (4) & 0.843 (0.040) & 0.709 (0.069) & 0.637 (0.077) & 0.553 (0.062) \\
    Subbotin (6) & 0.751 (0.055) & 0.645 (0.068) & 0.579 (0.063) & 0.545 (0.078) \\
    Subbotin (8) & 0.700 (0.064) & 0.562 (0.075) & 0.544 (0.076) & 0.522 (0.079) \\
    \hline
    Gaussian NS & 0.698 (0.052) & 0.541 (0.045) & 0.464 (0.061) & 0.469 (0.046) \\
    Gaussian Glasso & 0.677 (0.032) & 0.534 (0.045) & 0.456 (0.063) & 0.444 (0.055) \\
    \hline
    Quantile (0.5) & $<$ 0.001 ($<$ 0.001) & $<$ 0.001 ($<$ 0.001) & $<$ 0.001 ($<$ 0.001) & $<$ 0.001 ($<$ 0.001) \\
    Quantile (0.9) & 0.451 (0.044) & 0.324 (0.056) & 0.297 (0.052) & 0.169 (0.050) \\
    Quantile (0.99) & 0.688 (0.051) & 0.435 (0.054) & 0.444 (0.043) & 0.335 (0.044) \\
    \hline
    Copula (10) & \textbf{0.885} (0.042) & \textbf{0.751} (0.053) & \textbf{0.682} (0.066) & \textbf{0.616} (0.060) \\
    Copula (20) & 0.813 (0.053) & 0.657 (0.054) & 0.611 (0.057) & 0.529 (0.060) \\
    Copula (30) & 0.749 (0.059) & 0.612 (0.060) & 0.583 (0.051) & 0.502 (0.053) \\
    \hline
    \hline
    \textbf{Model}  & $\boldsymbol{n = 1000, p = 50}$ & $\boldsymbol{n = 2000, p = 50}$ & $\boldsymbol{n = 5000, p = 50}$ & $\boldsymbol{n = 10000, p = 50}$  \\
    \hline
    Subbotin (4) & 0.662 (0.070) & 0.687 (0.063) & 0.709 (0.069) & 0.712 (0.045) \\
    Subbotin (6) & 0.633 (0.069) & 0.636 (0.075) & 0.645 (0.068) & 0.651 (0.051) \\
    Subbotin (8) & 0.546 (0.066) & 0.558 (0.066) & 0.562 (0.075) & 0.607 (0.050) \\
    \hline
    Gaussian NS & 0.524 (0.055) & 0.537 (0.042) & 0.541 (0.045) & 0.572 (0.033) \\
    Gaussian Glasso & 0.525 (0.036) & 0.542 (0.043) & 0.534 (0.045) & 0.555 (0.031) \\
    \hline
    Quantile (0.5) & $<$ 0.001 ($<$ 0.001) & $<$ 0.001 ($<$ 0.001) & $<$ 0.001 ($<$ 0.001) & $<$ 0.001 ($<$ 0.001) \\
    Quantile (0.9) & 0.335 (0.059) & 0.316 (0.053) & 0.324 (0.056) & 0.339 (0.045) \\
    Quantile (0.99) & 0.414 (0.054) & 0.433 (0.052) & 0.435 (0.054) & 0.462 (0.039) \\
    \hline
    Copula (10) & \textbf{0.710} (0.057) & \textbf{0.717} (0.060) & \textbf{0.721} (0.063) & \textbf{0.751} (0.055) \\
    Copula (20) & 0.615 (0.053) & 0.631 (0.051) & 0.657 (0.054) & 0.670 (0.059) \\
    Copula (30) & 0.588 (0.055) & 0.590 (0.053) & 0.612 (0.060) & 0.618 (0.053) \\
    \hline
\end{tabular}
\end{footnotesize}
\end{table}

Here, we study the performance of our model on data generated from a multivariate block maxima model. We follow the procedure used by \citep{hy1} to analyze the block maxima copula model in order to generate the extreme value observations. We first generate data from a multivariate Gaussian distribution and then convert the marginal distributions of each variable to a standard GEV distribution with mean 5 using a copula transformation. For our simulations, we generate data from the multivariate Gaussian distribution with a sparse dependency structure, which we also use as the basis for comparison with the selected edges from each of the graphical models. The copularized GEV data is used as the maximal observations for blocks of size 10. The non-extreme observations are generated from a truncated multivariate Gaussian distribution with a covariance matrix containing diagonal entries of 1 and weak correlations between all pairs of features. We also show results for these simulation studies using data-driven stability selection hyperparameter tuning in Section A.2 in the Appendix.

In Table \ref{tab:bm}, we show the results for $p = 25, 50, 100, $ and $200$ with $n = 5000$, as well as for $n = 1000, 2000, 5000,$ and $10000$ with $p = 50$. In general, the block maxima copula graphical model with the correct block size performs the best in these simulations; this follows what we would expect, since the copula block maxima model is also the generative distribution for the simulated data. The Subbotin graphical model only performs slightly worse than the block maxima copula graphical model, with the best choice at $\nu = 4$. On the other hand, the Gaussian and quantile graphical models have substantially lower F1 scores compared to the Subbotin and block maxima copula models. Additionally, when the chosen block size is incorrect, the block maxima copula model is also not as accurate when compared to the Subbotin graphical model when $\nu = 4$. These results are encouraging, as they show that the Subbotin graphical model performs nearly as well as the true generative model and better than any of the non-correctly specified models in terms of edge selection with respect to the underlying conditional dependency graph for extreme value observations. We also observe that, when changing the number of observations and features, the Subbotin graphical model is somewhat less sensitive to smaller sample sizes relative to the number of features compared to the quantile and block maxima copula graphical models.

\subsection{Peaks-Over-Threshold Extreme Value Distribution} \label{sub:pot}

\begin{table}[t]
\caption{Average F1 scores (std. dev.) for the peaks-over-threshold simulations with oracle tuning, averaged over 50 replicates. Best performing methods are boldfaced. The Subbotin graphical model performs best out of all comparison methods, though the best choice of $\nu$ varies.}
\label{tab:pot}
\centering
\begin{footnotesize}
\begin{tabular}{|l|r|r|r|r|}
    \hline
    \textbf{Model}  & $\boldsymbol{n = 10000, p = 25}$ & $\boldsymbol{n = 10000, p = 50}$ & $\boldsymbol{n = 10000, p = 75}$ & $\boldsymbol{n = 10000, p = 100}$ \\
    \hline
    Subbotin (4) & 0.819 (0.025) & 0.765 (0.039) & 0.723 (0.040) & \textbf{0.684} (0.042) \\
    Subbotin (6) & 0.837 (0.026) & 0.792 (0.041) & \textbf{0.745} (0.045) & 0.665 (0.048) \\
    Subbotin (8) & \textbf{0.852} (0.032) & \textbf{0.804} (0.043) & 0.731 (0.044) & 0.643 (0.049) \\
    \hline
    Gaussian NS & 0.662 (0.037) & 0.571 (0.037) & 0.529 (0.033) & 0.461 (0.039) \\
    Gaussian Glasso & 0.681 (0.037) & 0.560 (0.035) & 0.510 (0.034) & 0.457 (0.035) \\
    \hline
    Quantile (0.5) & $<$ 0.001 ($<$ 0.001) & $<$ 0.001 ($<$ 0.001) & $<$ 0.001 ($<$ 0.001) & $<$ 0.001 ($<$ 0.001) \\
    Quantile (0.9) & 0.663 (0.030) & 0.602 (0.035) & 0.545 (0.036) & 0.486 (0.041) \\
    Quantile (0.99) & 0.732 (0.032) & 0.689 (0.038) & 0.642 (0.034) & 0.519 (0.042) \\
    \hline
    Copula (10) & 0.163 (0.017) & 0.087 (0.019) & 0.066 (0.010) & 0.065 (0.008) \\
    Copula (20) & 0.355 (0.030) & 0.204 (0.015) & 0.190 (0.013) & 0.167 (0.018) \\
    Copula (30) & 0.326 (0.025) & 0.130 (0.020) & 0.112 (0.019) & 0.114 (0.012) \\
    \hline
    \hline
    \textbf{Model}  & $\boldsymbol{n = 5000, p = 50}$ & $\boldsymbol{n = 7500, p = 50}$ & $\boldsymbol{n = 10000, p = 50}$ & $\boldsymbol{n = 12500, p = 50}$  \\
    \hline
    Subbotin (4) & 0.652 (0.050) & 0.745 (0.046) & 0.765 (0.039) & 0.801 (0.032) \\
    Subbotin (6) & \textbf{0.653} (0.053) & \textbf{0.754} (0.048) & 0.792 (0.041) & 0.803 (0.035) \\
    Subbotin (8) & 0.608 (0.055) & 0.687 (0.045) & \textbf{0.804} (0.043) & \textbf{0.824} (0.037) \\
    \hline
    Gaussian NS & 0.479 (0.042) & 0.523 (0.034) & 0.571 (0.037) & 0.673 (0.030) \\
    Gaussian Glasso & 0.451 (0.040) & 0.517 (0.040) & 0.560 (0.035) & 0.661 (0.029) \\
    \hline
    Quantile (0.5) & $<$ 0.001 ($<$ 0.001) & $<$ 0.001 ($<$ 0.001) & $<$ 0.001 ($<$ 0.001) & $<$ 0.001 ($<$ 0.001) \\
    Quantile (0.9) & 0.440 (0.038) & 0.531 (0.039) & 0.602 (0.035) & 0.685 (0.027) \\
    Quantile (0.99) & 0.544 (0.042) & 0.562 (0.040) & 0.689 (0.038) & 0.732 (0.035) \\
    \hline
    Copula (10) & 0.082 (0.019) & 0.079 (0.015) & 0.087 (0.019) & 0.106 (0.023) \\
    Copula (20) & 0.131 (0.020) & 0.188 (0.021) & 0.204 (0.015) & 0.255 (0.022) \\
    Copula (30) & 0.045 (0.018) & 0.109 (0.019) & 0.130 (0.020) & 0.167 (0.024) \\
    \hline
\end{tabular}
\end{footnotesize}
\end{table}

We now compare the Subbotin and other graphical model methods using data generated from a peaks-over-threshold model. To do this, we utilize a two-step process. We first model occurrences of extreme values as a multivariate Hawkes process \citep{rz1}, which allows us to simulate correlated occurrences of extreme value observations between different features at varied intervals. We impose a sparse pairwise dependency structure on the Hawkes process in order to mirror a graphical model distribution. The magnitude of the extreme values are then determined by adding data simulated from a standard GEV distribution to a pre-chosen threshold of 10. Non-extreme observations are generated from a multivariate Gaussian distribution with a covariance matrix containing diagonal entries of 1 and weak correlations between all pairs of features; we also truncate these observations to ensure that they are less than the chosen threshold. We compare the selected edges from each of the graphical models methods to the underlying dependency structure of the multivariate Hawkes process in the first simulation step. We also show results for these simulation studies using data-driven stability selection hyperparameter tuning in Section A.3 in the Appendix.

Table \ref{tab:pot} shows the average F1 scores of the graph estimates for $p = 25, 50, 75, $ and $100$ with $n = 10000$, as well as for $n = 5000, 7500, 10000,$ and $12500$ with $p = 50$. Here, the Subbotin graphical model performs better than all of the other models we compare. In particular, $\nu = 8$ has higher F1 scores when the ratio of observations to features is larger, while $\nu = 4$ does better when the number of observations is proportionally smaller. On the other hand, the block maxima copula model performs substantially worse than any of the other models. This is explained by the distribution of extreme value observations over the full data set; since these do not occur at regular intervals, many of the block maxima are likely not actual extreme value observations and thus not useful for finding the conditional dependencies between extreme values. The Gaussian and quantile graphical models also perform decently, but are not as accurate as the Subbotin graphical model with any $\nu$ of comparison.

\section{Functional Neuronal Connectivity Case Study}

\begin{figure}[t]
    \begin{subfigure}[t]{0.4\linewidth}
    \centering
        \includegraphics[width=\linewidth]{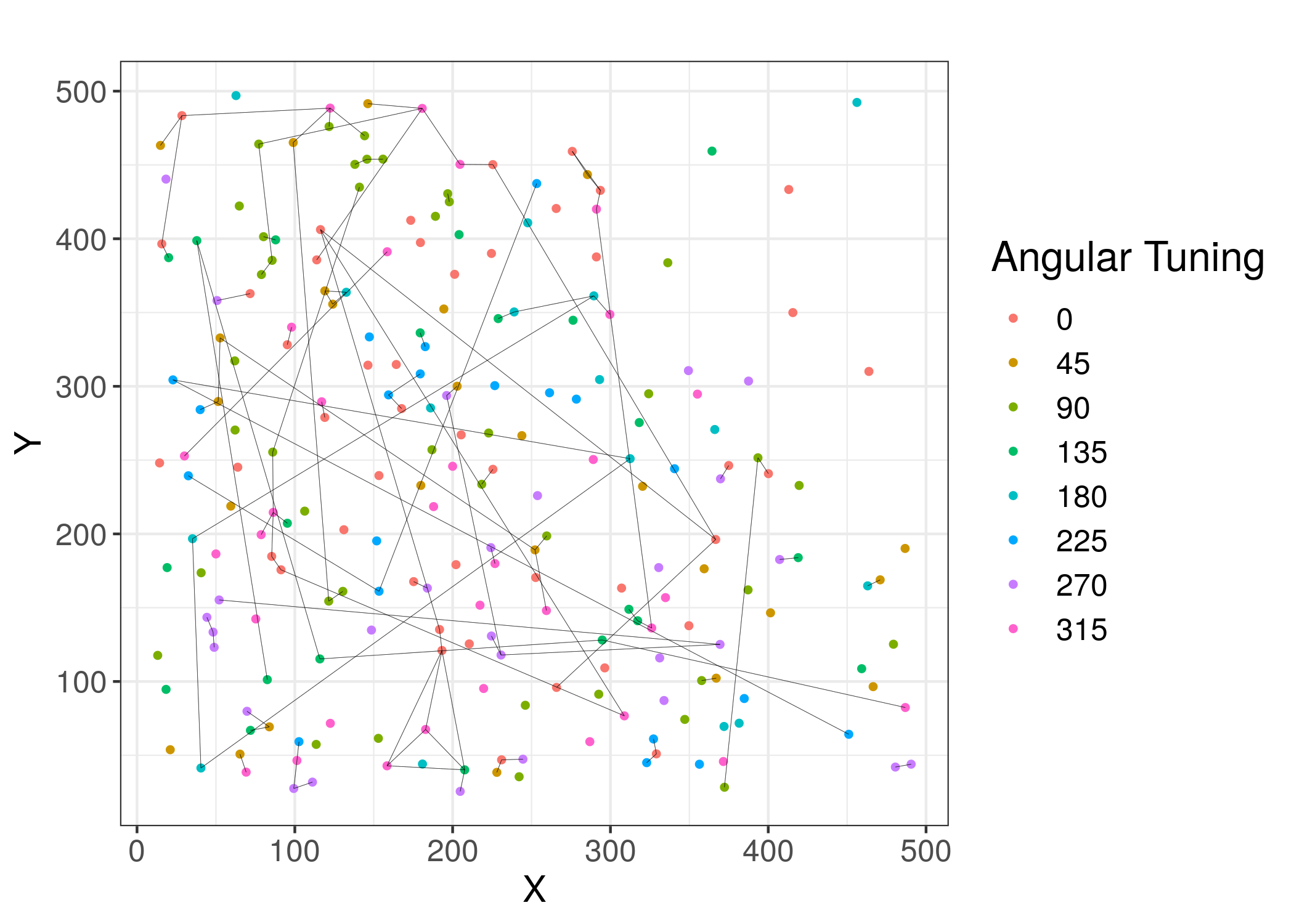}
        \caption{Gaussian graphical model.}
    \label{fig:nsg}
    \end{subfigure}
    \begin{subfigure}[t]{0.4\linewidth}
    \centering
        \includegraphics[width=\linewidth]{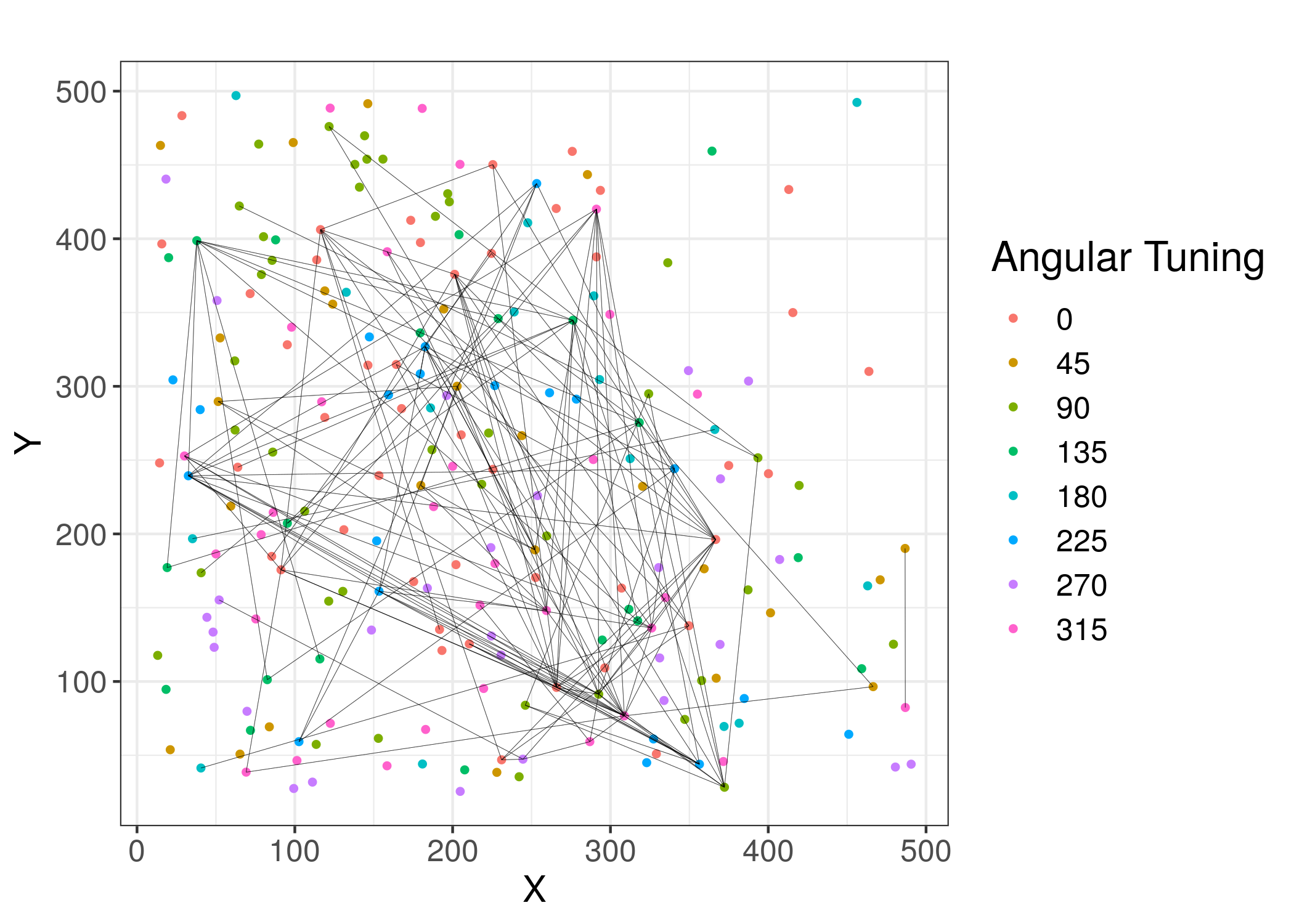}
        \caption{Subbotin graphical model.}
    \label{fig:nsg2}
    \end{subfigure}
    \begin{subfigure}[t]{0.4\linewidth}
    \centering
        \includegraphics[width=\linewidth]{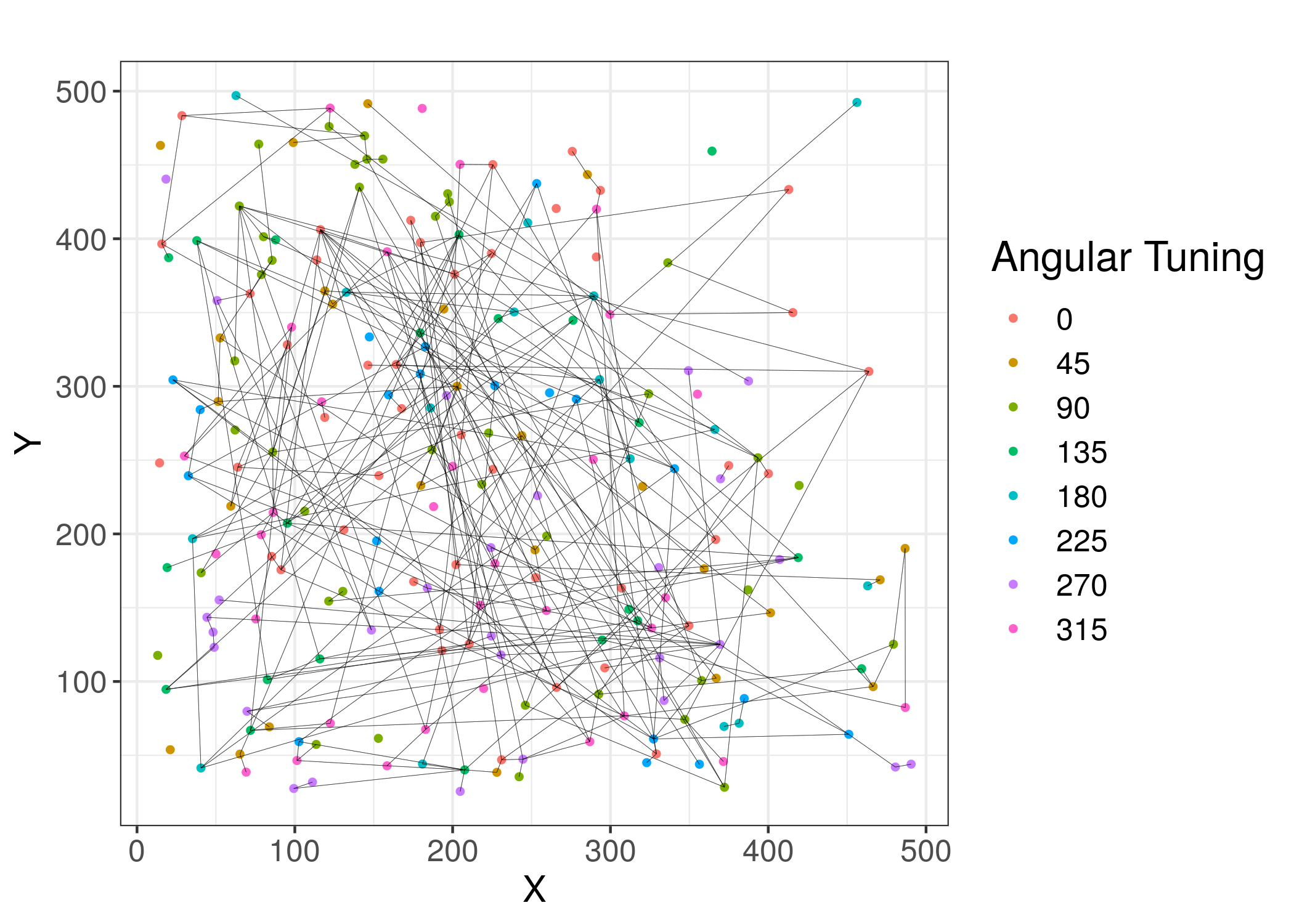}
        \caption{Quantile graphical model ($q = 0.99$)}
    \label{fig:nsg3}
    \end{subfigure}
    \begin{subfigure}[t]{0.4\linewidth}
    \centering
        \includegraphics[width=\linewidth]{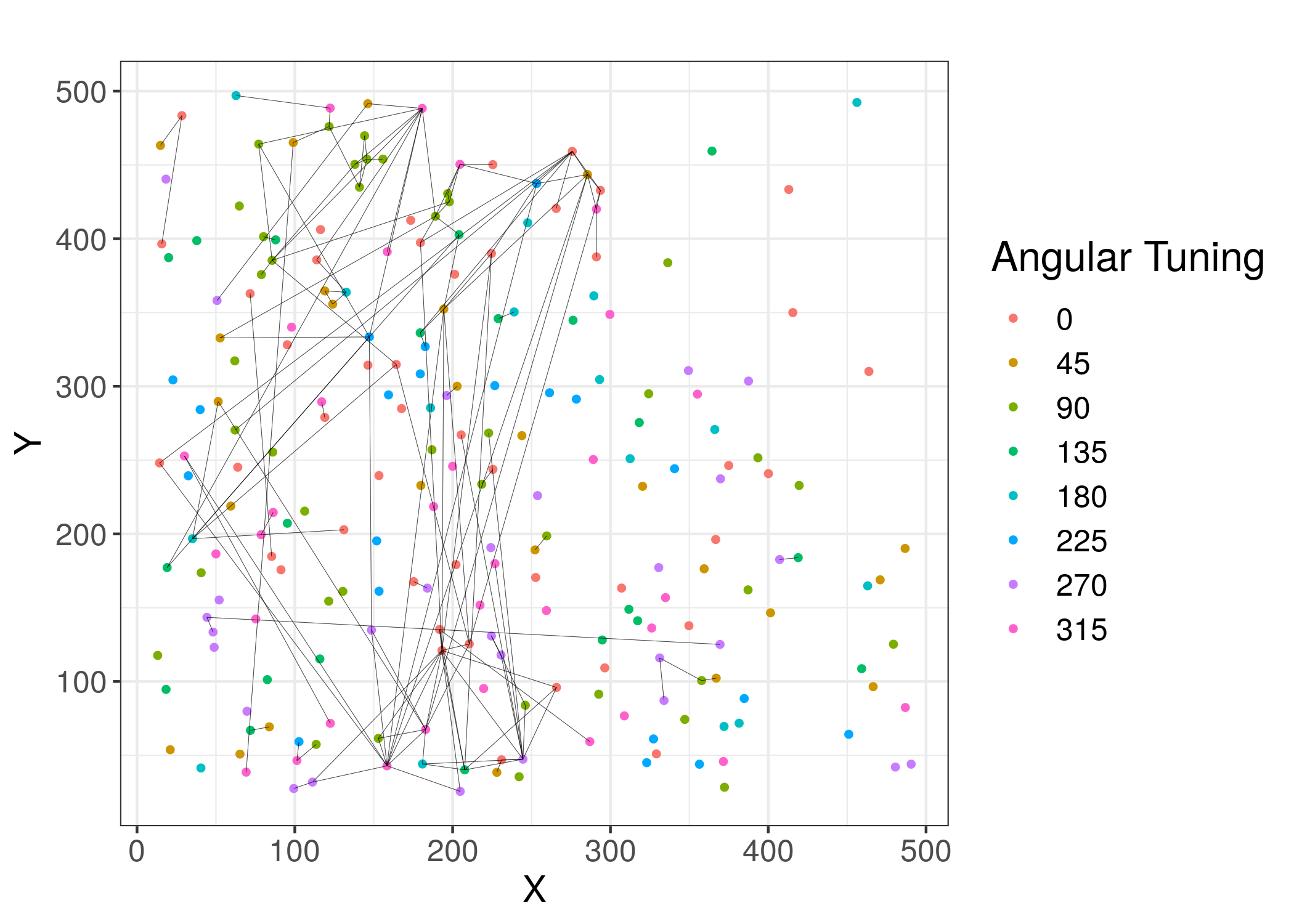}
        \caption{Copula graphical model}
    \label{fig:nsg4}
    \end{subfigure}
    
    \caption{Comparison of the functional connectivity estimates from the graphical modeling techniques. Neurons are plotted with respect to their spatial locations and colored by their angular tuning category.}
    \label{fig:g1}
\end{figure}

Lastly, we investigate the applicability of the Subbotin graphical model to estimating functional neuronal connectivity networks for real-world calcium imaging data. We use a data set originally from the Allen Brain Atlas \citep{aba1} which has fluorescence traces for 227 neurons in the visual cortex of the brain of a mouse at over 100000 time point observations; for this particular case study, we examine a subset of 18000 time points from this data set during a period of controlled drifting grating visual stimulus activity. Our goal is to estimate the intrinsic functional connectivity of the neurons, i.e. correlations between their contemporaneous firing activities. We compare the estimated functional neuronal connectivity graphs from the Subbotin graphical model to those from previous approaches from the graphical model literature, namely the Gaussian, quantile, and block maxima extreme graphical models, as well as several of the functional connectivity models introduced in Section \ref{sec:intro}, including the Hawkes model \citep{me2}, VAR model \citep{mk1}, transfer entropy model \citep{mg1}, and linear-nonlinear Poisson model \citep{ip1}. For the block maxima model, we look at block sizes of 15, 30, and 45, while for the quantile graphical model, we look at the 0.5, 0.9, 0.99, and 0.999 percentiles. Again, we do not include the multivariate Pareto graphical model here because of the restrictions on the model. Hyperparameter selection for the graphical model methods are performed using stability selection with a threshold of 0.95. Edge selection is performed for the Hawkes, VAR, and transfer entropy models by selecting the highest magnitude parameter estimates. An ANOVA comparison with a BIC criterion is used to select edges for the linear-nonlinear Poisson model. In the cases where there is no data-driven process for choosing graph sparsity for the functional connectivity models, we select the same number of edges as selected by the Subbotin graphical model.

\begin{table}[t]
\caption{Proportion of edges from the estimated functional connectivity network by each model which link two neurons of the same tuning category.}
\label{tab:freq}
\medskip
\begin{footnotesize}
\begin{tabular}{|l|r|r|}
    \hline
    \textbf{Model} & Angular Tuning & Frequency Tuning\\
    \hline
    Subbotin ($\nu = 10$) & 0.578 & 0.642 \\
    \hline
    Gaussian NS & 0.417 & 0.426 \\
    Gaussian Glasso & 0.429 & 0.449 \\
    \hline
    Quantile (0.5) & 0.127 & 0.256\\
    Quantile (0.9) & 0.358 & 0.349 \\
    Quantile (0.99) & 0.494 & 0.491 \\
    Quantile (0.999) & 0.446 & 0.485 \\
    \hline
    Copula (15) & 0.207 &  0.289 \\
    Copula (30) & 0.292 &  0.262 \\
    Copula (45) & 0.253 &  0.280 \\
    \hline
    Hawkes & 0.140 & 0.335 \\
    Transfer Entropy & 0.243 & 0.193 \\
    VAR & 0.112 & 0.387 \\
    Linear-Nonlinear & 0.460 & 0.434 \\
    \hline
\end{tabular}
\end{footnotesize}
\end{table}

We first analyze the selected edges from each model. In Figure \ref{fig:g1}, we show the functional neuronal connectivity estimates from the different graphical model methods, with each graph estimate plotted with respect to the spatial locations of the neurons. Notably, the Subbotin graphical model produces a functional connectivity network which is structurally small-world with a few hub neurons, which matches what has been previously proposed about functional connectivity in the brain \citep{os1, cs1, cp1}; we do not see these architectural features nearly as much for the estimates from the other models. In Table \ref{tab:freq}, we compare the selected edges from each model with respect to the orientation and frequency tunings of the pairs of neurons in the edge set. It is hypothesized in the neuroscience literature that neurons are tuned to fire when presented with particular stimuli, and that these neurons are more likely to be connected functionally \citep{tune}; thus, we would expect that the estimated functional connectivity network should reflect the visual tuning of the observed neurons, meaning that a substantial proportion of the function connections should be between neurons with the same tuning. For this data set, the neuron tuning with respect to visual angles and frequencies have been estimated a priori. There are 8 categories of angular tuning and 3 categories of frequency tuning, which correspond to the different visual stimulus shown to the mouse during the recording. In Table \ref{tab:freq}, we show the proportions of edges with matching tuning categories from all comparison methods. We see that the functional connections selected by the Subbotin graphical model are most likely to be associated with pairs of neurons with the same category of neuron tuning, both in terms of frequency and angular visual stimuli.  Of the other models compared in Table \ref{tab:freq}, the linear-nonlinear and Gaussian graphical models perform moderately worse, while the rest of the models are very poor with regards to estimating functional connections that match neurons with the same tuning.

\begin{figure}[t]
    \centering
    \begin{subfigure}[t]{0.49\linewidth}
        \includegraphics[width=\linewidth]{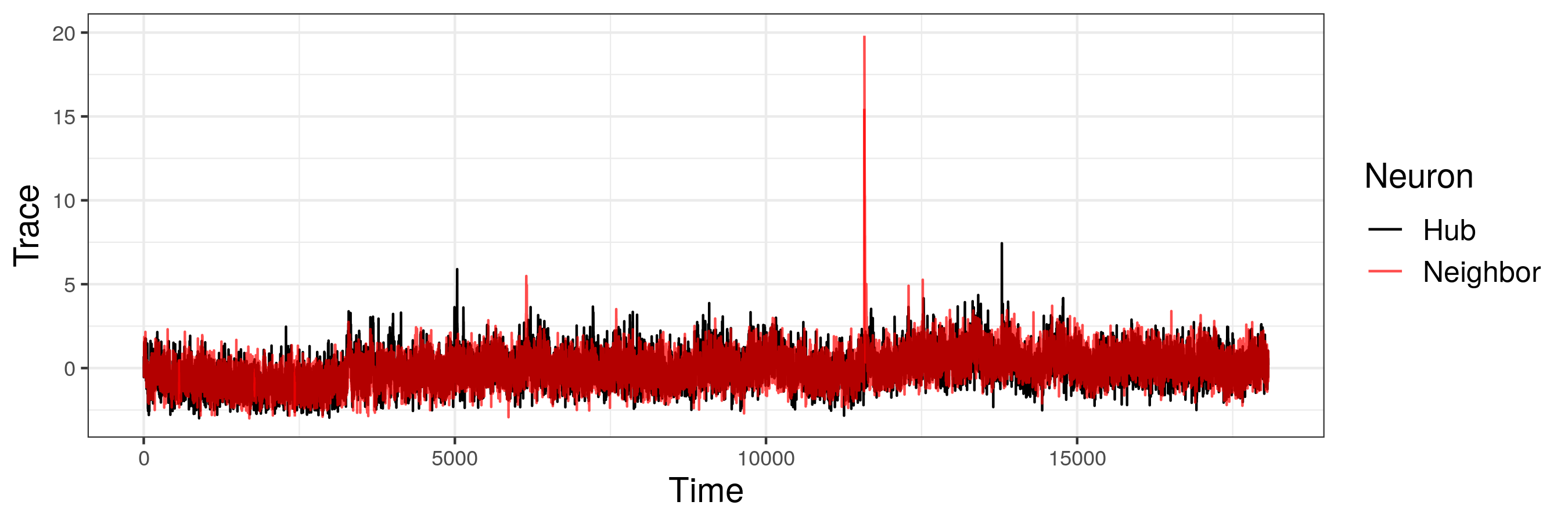}
        \includegraphics[width=\linewidth]{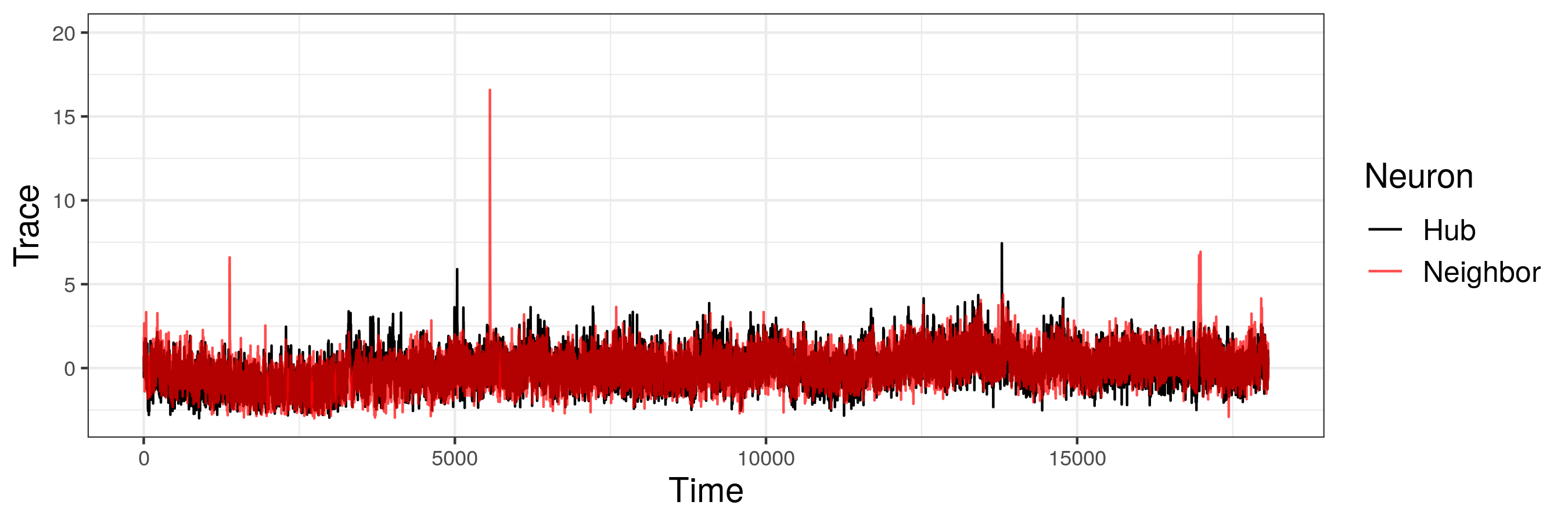}
    \caption{Gaussian graph hub neuron.}
    \label{fig:ns}
    \end{subfigure} %
    \begin{subfigure}[t]{0.49\linewidth}
        \includegraphics[width=\linewidth]{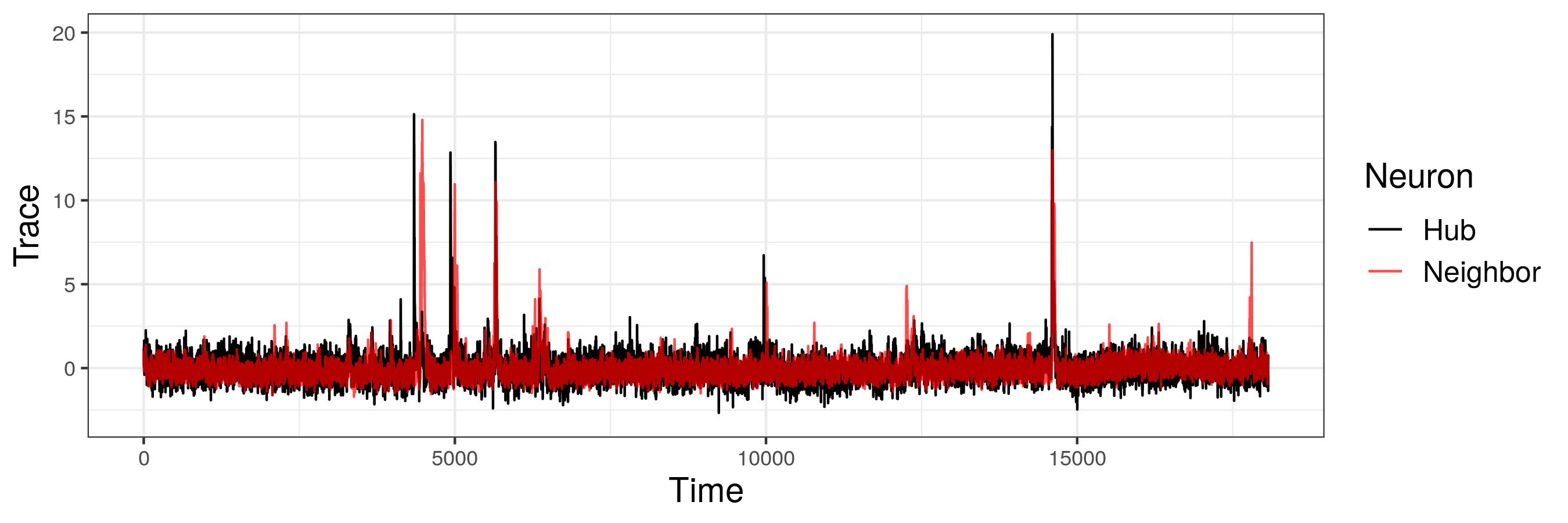}
        \includegraphics[width=\linewidth]{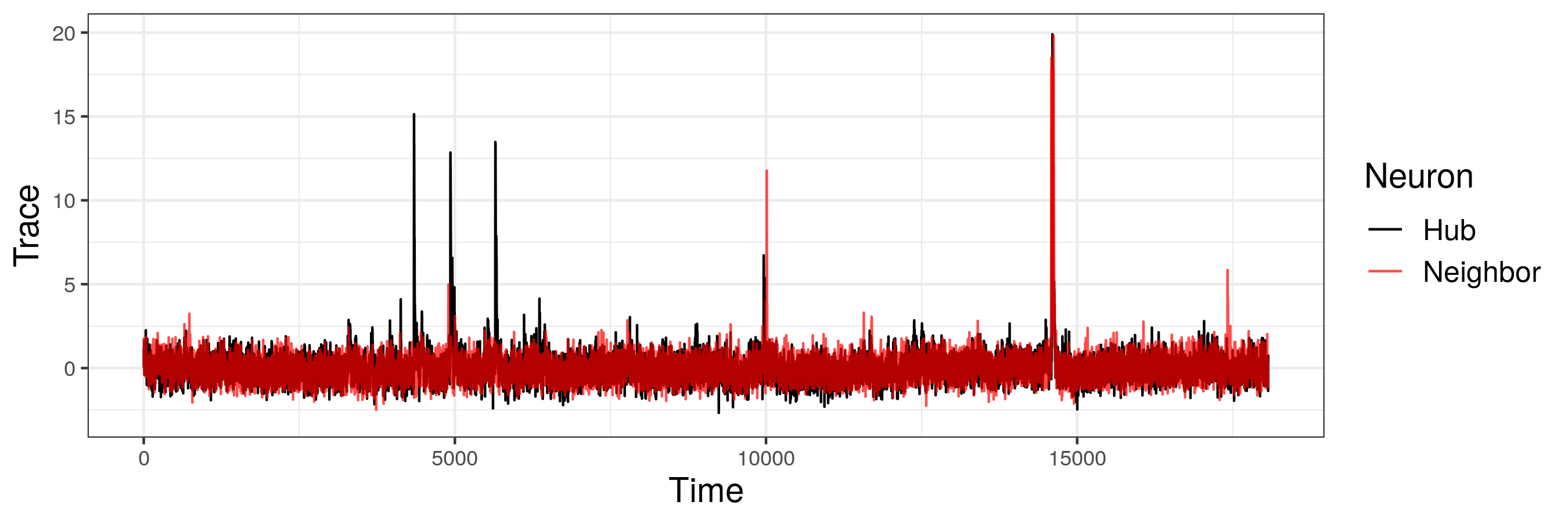}
        \caption{Extreme graph hub neuron.}
        \label{fig:ens}
    \end{subfigure}
    \caption{Trace of different high degree hub neurons (black) vs. traces of selected edge neighbors (red) in functional connectivity estimates from the Gaussian and Subbotin graphical models.}
    \label{fig:hub}
\end{figure}

We then look at the selected edges for the largest degree hub neurons in the estimated functional connectivity graphs from the Subbotin graphical model and from the Gaussian graphical model. In Figure \ref{fig:hub}, we show the fluorescence traces for each respective hub neuron and its neighbors plotted on top of one another. The fluorescence traces of the hub neuron of the Gaussian graphical model and its neighbors, as seen in in Figure \ref{fig:ns}, display a similar baseline mean shift pattern across time, possibly caused by artifacts of the data collection methodology. However, there are clearly no relationships between the hub neuron and any of the selected neighbors with respect to any extreme value spikes in the fluorescence traces, and it does not appear that the selected hub neuron is particularly active at all during the recorded time period. On the other hand, the hub neuron of the Subbotin graphical model clearly fire simultaneously with its selected graph neighbors more than once during the experiment, as seen in in Figure \ref{fig:ens}. Thus, the edges from the Subbotin graphical model appear to fit the definition of intrinsic functional connectivity much more closely than those from the Gaussian graphical model.

\begin{figure}[t]
    \centering
    \begin{subfigure}[t]{0.49\linewidth}
        \includegraphics[width=\linewidth]{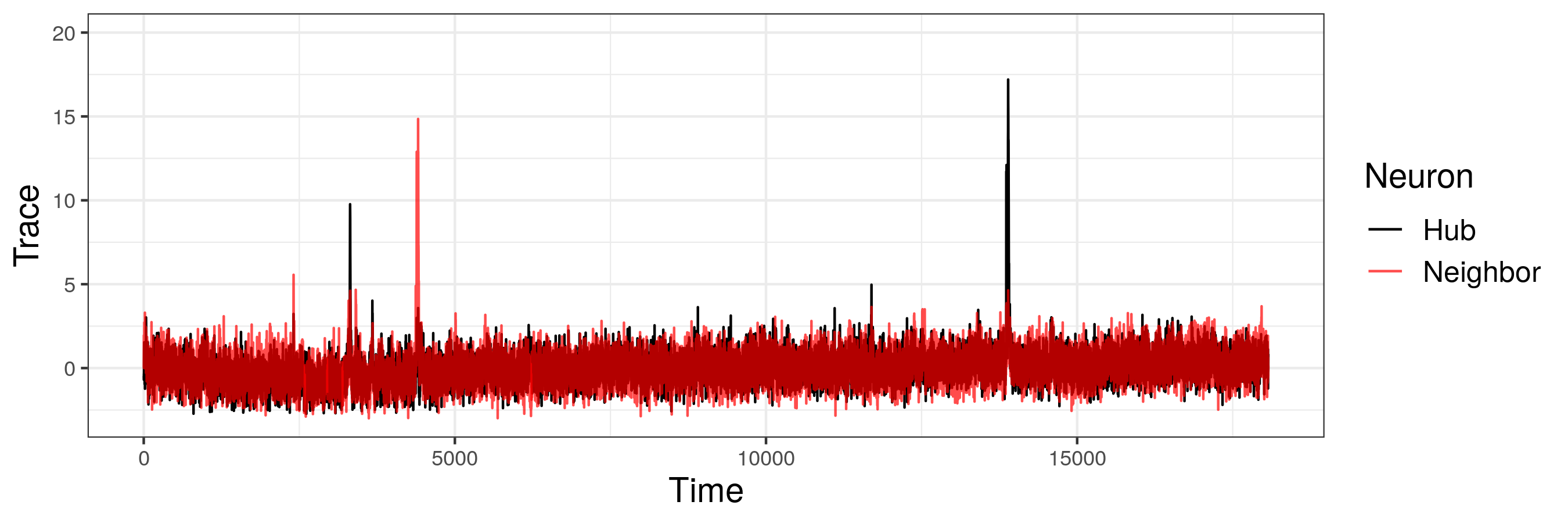}
        \includegraphics[width=\linewidth]{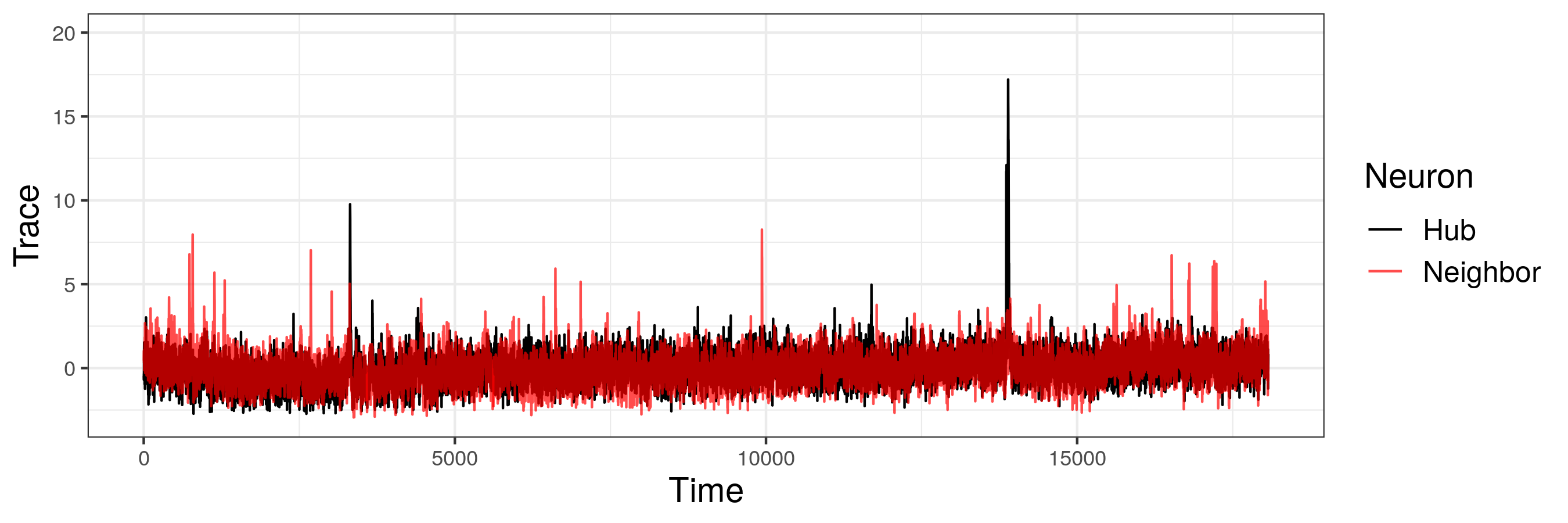}
    \caption{Gaussian graph connections.}
    \label{fig:nss}
    \end{subfigure} %
    \begin{subfigure}[t]{0.49\linewidth}
        \includegraphics[width=\linewidth]{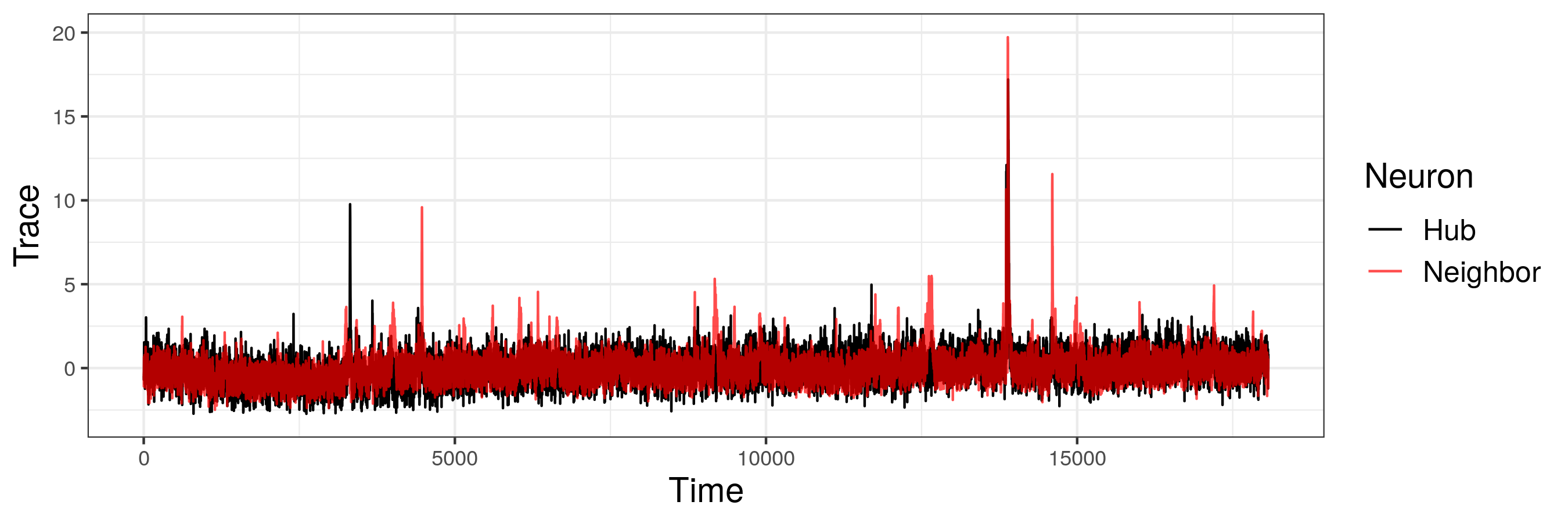}
        \includegraphics[width=\linewidth]{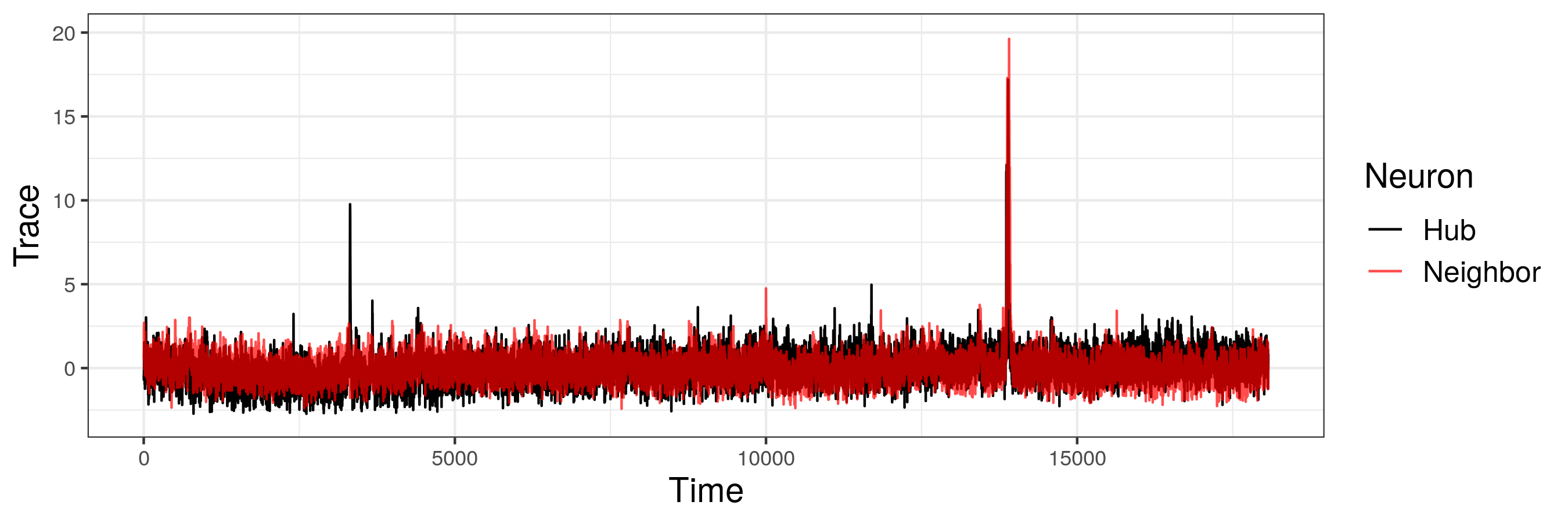}
        \caption{Extreme graph connections.}
        \label{fig:enss}
    \end{subfigure}
    \caption{Trace of single particular neuron (black) vs. trace of selected edge neighbors (red) in functional connectivity estimates from the Gaussian and Subbotin graphical models.}
    \label{fig:same}
\end{figure}

We also show the differences in the edges selected that involve the same neuron in the graph estimates from the Subbotin and Gaussian graphical models in Figure \ref{fig:same}. The neuron we highlight here has several edge neighbors in the estimated networks for both the Gaussian and Subbotin graphical models, but the neighbors selected by each method are different. We can see from Figure \ref{fig:nss} that the neighbors in the graph estimated by the Gaussian graphical model do not appear to share any similar spiking patterns with the highlighted neuron. On the other hand, as seen from Figure \ref{fig:enss}, the edge neighbors as selected by the Subbotin graphical model for the exact same neuron seem to spike at the same time at one particular time point during the recording. From this example, the Subbotin graphical model seems to do a better job of finding neurons which share similar spiking patters, as is the goal when estimating functional neuronal connectivity networks.

\section{Discussion}

In this paper, we have introduced the Subbotin graphical model for estimating networks for data in which the primary focus of the analysis is on finding the conditional dependency structure with respect to the extreme value observations. Our model is constructed by defining the node-wise conditional distribution of the graphical model as a Subbotin distribution with $\nu > 2$, which increases the relative weight of extreme values in the estimation process compared to the Gaussian graphical model. Our proposed model permits simultaneous parameter estimation and automatic sparse model selection through the probabilistic graphical modeling framework, which many current functional connectivity models do not allow. The Subbotin graphical model has several advantages over other graphical modeling methods used for this type of analysis as well. Unlike methods in the existing literature, the Subbotin graphical model does not require any data pre-processing with binning or thresholding to identify extreme values before model estimation in order to produce meaningful results, instead smoothly increasing the relative weight of extreme value observations based on their magnitude. It also does not require an input prior like the multivariate Pareto graphical model or a particular selection of a quantile for which to find dependencies. Through simulation studies and a real-world example, we have shown the potential usefulness of the Subbotin graphical model to analyzing data with extreme value observations. In particular, we showed the suitability of our model to neuroscientific applications, specifically toward estimating intrinsic functional neuronal connectivity graphs for calcium imaging data.

There are several possible areas of research for the Subbotin graphical model that can be investigated in future works. From a methodological standpoint, we have used ordinary bootstrapping sampling with our stability selection method in order to select the hyperparameter values for the individual node-wise regressions. However, this type of procedure may not be the most appropriate when the main focus of the analysis is on the rare extreme value observations, and there may be better schemes that could be implemented for this part of the algorithm. Also, like with the Gaussian graphical model, there are many potential extensions to the Subbotin graphical model that could be developed to account for different experimental conditions or possible effects. For example, in neuroscience, common implementations of graphical model methods include latent variable adjustments to adjust for the potential impact of unmeasured brain neuron activity or stimuli, covariate adjustment methods, and changepoint methods to create smoothly varying functional neuronal connectivity estimates under different experimental stimuli conditions. 

We also note here that the Subbotin graphical model that we have developed is in this paper is based on an assumption of independent observations, which we chose for this particular application because of our stated goal of finding intrinsic functional neuronal connectivity, i.e. only contemporaneously correlated neuronal activity. While there are potential autoregressive effects present in the fluorescence traces produced by calcium imaging  for which the Subbotin graphical model described in this work does not account for, previous literature shows that the methodology described in this paper can still be applied successfully in the presence of these effects. Specifically, in \citep{sb1}, it is shown that Lasso-based sparse estimators, such as the one used for the Subbotin graphical model, are still model selection consistent in this case, though with the a larger theoretical sample complexity. Because the number of observed time points in calcium imaging data sets is often large relative to the number of neurons that are in an individual recording, we believe our method will work well even in the presence of these autocorrelations. However, in many examples of data sets for which our method may be relevant, it may be assumed that the observed features follow some type of time series model which need to be accounted for \citep{ep1}, or the problem of interest may involve directed networks across time lags. Thus, a methodological extension for time series data could be a pertinent area of investigation for future work.

The Subbotin graphical model has the potential to be explored further as well for different applied research problems. Within the realm of neuroscience, our model could be used to study neuronal functional connectivity networks for calcium imaging data on a much larger scope than the one considered in this paper, with possibly up to tens of thousands of in vivo simultaneously recorded neurons. These studies can help provide new insights into the how individual neurons in the brain are organized functionally on a large-scale basis. Also, as mentioned in Section \ref{sec:intro}, one of our future research goals is to apply our method in order to explore the relationship between functional and structural connectivity in the brain. Some particular interesting studies could involve comparing how closely estimated functional connectivity network from both our model as well as the other common functional connectivity in the literature match the brain's physical structure and synaptic connectivity, or attempting to characterize how certain patterns in stimuli or activity may be associated with differences between the two. Additionally, while the main focus of the paper has involved relating our method to neuroscience experiments, there are several other fields in which the Subbotin graphical model could be utilized, including climatology, signal processing, and finance. In conclusion, our work has created a new single-step graphical model method which could provide a more accurate method for estimating scientifically meaningful networks for data in which the relevant information in the data involves the extreme value observations.

\begin{funding}

The authors were supported by NSF NeuroNex-1707400, NIH 1R01GM140468, and NSF DMS-2210837.

\end{funding}

\begin{supplement} \stitle{Supplementary Information}

\sdescription{Results of data-driven edge selection for the simulation studies. Proofs and derivations for Theorems 1, 2, and 3.}

\end{supplement}

\newpage

\appendix

\section{Data-Driven Selection for Simulation Studies}

\begin{table}[b]
\caption{Average F1 scores (std. devs) for multivariate Subbotin graphical model distribution simulations with data-driven edge selection, averaged over 50 replicates. Best performing methods are boldfaced.}
\label{tab:mvgn}
\centering
\begin{footnotesize}
\begin{tabular}{|l|r|r|r|r|}
    \hline
    \textbf{Model}  & $\textbf{n = 500, p = 100}$ & $\textbf{n = 1000, p = 100}$ & $\textbf{n = 2000, p = 100}$ & $\textbf{n = 4000, p = 100}$ \\
    \hline
    Subbotin (4) & 0.522 (0.073) & 0.598 (0.065) & 0.722 (0.070) & 0.745 (0.065) \\
    Subbotin (6) & 0.531 (0.072) & 0.648 (0.070) & 0.756 (0.066) & 0.788 (0.064) \\
    Subbotin (8) & \textbf{0.579} (0.074) & \textbf{0.661} (0.070) & \textbf{0.779} (0.071) & \textbf{0.801} (0.062) \\
    \hline
    Gaussian NS & 0.436 (0.065) & 0.547 (0.059) & 0.694 (0.060) & 0.723 (0.055) \\
    Gaussian Glasso & 0.445 (0.067) & 0.555 (0.061) & 0.705 (0.058) & 0.712 (0.056) \\
    \hline
    Quantile (0.5) & 0.424 (0.053) & 0.498 (0.054) & 0.601 (0.057) & 0.665 (0.058) \\
    Quantile (0.9) & 0.188 (0.050) & 0.207 (0.056) & 0.213 (0.055) & 0.254 (0.051) \\
    Quantile (0.99) & $<$ 0.001 ($<$ 0.001) & $<$ 0.001 ($<$ 0.001) & $<$ 0.001 ($<$ 0.001) & $<$ 0.001 ($<$ 0.001) \\
    \hline
    Copula (10) & 0.319 (0.056) & 0.471 (0.054) & 0.554 (0.055) & 0.598 (0.050) \\
    Copula (20) & 0.304 (0.057) & 0.465 (0.053) & 0.536 (0.059) & 0.562 (0.056) \\
    Copula (30) & 0.296 (0.054) & 0.439 (0.051) & 0.521 (0.055) & 0.554 (0.054) \\
    \hline
    \hline
    \textbf{Model}  & $\textbf{n = 2000, p = 30}$ & $\textbf{n = 2000, p = 100}$ & $\textbf{n = 2000, p = 300}$ & $\textbf{n = 2000, p = 500}$ \\
    \hline
    Subbotin (4) & 0.735 (0.065) & 0.722 (0.070) & 0.646 (0.068) & 0.463 (0.063) \\
    Subbotin (6) & 0.788 (0.070) & 0.756 (0.066) & 0.651 (0.065) & 0.507 (0.064) \\
    Subbotin (8) & \textbf{0.831} (0.070) & \textbf{0.779} (0.071) & \textbf{0.668} (0.068) & \textbf{0.508} (0.069) \\
    \hline
    Gaussian NS & 0.755 (0.054) & 0.694 (0.060) & 0.552 (0.065) & 0.430 (0.066) \\
    Gaussian Glasso & 0.758 (0.057) & 0.705 (0.058) & 0.562 (0.068) & 0.436 (0.067) \\
    \hline
    Quantile (0.5) & 0.665 (0.062) & 0.601 (0.057) & 0.514 (0.063) & 0.417 (0.067) \\
    Quantile (0.9) & 0.278 (0.051) & 0.213 (0.055) & 0.191 (0.056) & 0.195 (0.053) \\
    Quantile (0.99) & $<$ 0.001 ($<$ 0.001) & $<$ 0.001 ($<$ 0.001) & $<$ 0.001 ($<$ 0.001) & $<$ 0.001 ($<$ 0.001) \\
    \hline
    Copula (10) & 0.632 (0.048) & 0.554 (0.055) & 0.489 (0.054) & 0.378 (0.053) \\
    Copula (20) & 0.624 (0.049) & 0.536 (0.059) & 0.477 (0.056) & 0.341 (0.051) \\
    Copula (30) & 0.625 (0.043) & 0.521 (0.055) & 0.442 (0.053) & 0.332 (0.050) \\
    \hline
    \hline
    \textbf{Model}  & $\boldsymbol{\nu = 4}$ & $\boldsymbol{\nu = 6}$ & \textbf{Chain Graph} & \textbf{Erdos-Renyi Graph} \\
    \hline
    Subbotin (4) & \textbf{0.821} (0.062) & 0.770 (0.070) & 0.647 (0.069) & 0.576 (0.073) \\
    Subbotin (6) & 0.794 (0.062) & \textbf{0.807} (0.065) & 0.655 (0.072) & 0.572 (0.070) \\
    Subbotin (8) & 0.742 (0.067) & 0.749 (0.062) & \textbf{0.691} (0.068) & \textbf{0.577} (0.078) \\
    \hline
    Gaussian NS & 0.808 (0.056) & 0.761 (0.062) & 0.626 (0.064) & 0.557 (0.063) \\
    Gaussian Glasso & 0.810 (0.048) & 0.769 (0.055) & 0.621 (0.052) & 0.564 (0.060) \\
    \hline
    Quantile (0.5) & 0.614 (0.056) & 0.605 (0.061) & 0.556 (0.067) & 0.534 (0.063)  \\
    Quantile (0.9) & 0.221 (0.049) & 0.230 (0.050) & 0.181 (0.053) & 0.188 (0.052)\\
    Quantile (0.99) & $<$ 0.001 ($<$ 0.001) & $<$ 0.001 ($<$ 0.001) & $<$ 0.001 ($<$ 0.001) & $<$ 0.001 ($<$ 0.001) \\
    \hline
    Copula (10) & 0.545 (0.050) & 0.550 (0.057) & 0.538 (0.052) & 0.473 (0.054) \\
    Copula (20) & 0.529 (0.052) & 0.542 (0.056) & 0.536 (0.056) & 0.458 (0.053) \\
    Copula (30) & 0.521 (0.048) & 0.520 (0.059) & 0.506 (0.055) & 0.436 (0.056) \\
    \hline
\end{tabular}
\end{footnotesize}
\end{table}

In this section, we present results for each of our three simulation studies, with data driven selection of hyperparameter values. We describe the specific procedure for hyperparameter selection in each subsection below.

\subsection{Subbotin Graphical Model Distribution}

Table \ref{tab:mvgn} shows the results from the Subbotin graphical model distribution simulations with data-driven model selection via stability selection. For each simulation set, we show the average and standard deviation of the F1 score for each model with respect to estimating the true edge set of the underlying graph of the data across 50 replications. We resample the original data with the same number of observations as the original data created using block bootstrapping with random block lengths, and we use a stability threshold of 0.95. Here, we only use stability selection to choose $\lambda$ for the Subbotin graphical model, instead of both $\nu$ and $\lambda$. Overall, we see that all methods perform slightly worse when compared to oracle sparsity tuning. The Subbotin graphical model with the correctly specified $\nu$ still performs best. However, the Gaussian graphical model sometimes outperforms the Subbotin graphical model with a misspecified $\nu$; this could potentially be caused by sensitivity of the model to outliers in the simulated data. The block maxima copula graphical model and quantile graphical models perform nearly as well as for oracle sparsity tuning, which could indicate that these models are relatively more robust for this type of data.

\begin{table}[t]
\caption{Average F1 scores (std. dev.) for the block maxima simulations with data-driven edge selection, averaged over 50 replicates. Best performing methods are boldfaced.}
\label{tab:bm}
\centering
\begin{footnotesize}
\begin{tabular}{|l|r|r|r|r|}
    \hline
    \textbf{Model}  & $\textbf{n = 5000, p = 25}$ & $\textbf{n = 5000, p = 50}$ & $\textbf{n = 5000, p = 100}$ & $\textbf{n = 5000, p = 200}$ \\
    \hline
    Subbotin & 0.783 (0.057) & 0.657 (0.068) & 0.611 (0.078) & 0.502 (0.079) \\
    \hline
    Gaussian NS & 0.624 (0.062) & 0.521 (0.064) & 0.455 (0.061) & 0.446 (0.065) \\
    Gaussian Glasso & 0.621 (0.058) & 0.517 (0.067) & 0.423 (0.063) & 0.415 (0.069) \\
    \hline
    Quantile (0.5) & $<$ 0.001 ($<$ 0.001) & $<$ 0.001 ($<$ 0.001) & $<$ 0.001 ($<$ 0.001) & $<$ 0.001 ($<$ 0.001) \\
    Quantile (0.9) & 0.423 (0.064) & 0.308 (0.060) & 0.301 (0.055) & 0.188 (0.057) \\
    Quantile (0.99) & 0.620 (0.078) & 0.397 (0.064) & 0.422 (0.066) & 0.325 (0.064) \\
    \hline
    Copula (10) & \textbf{0.808} (0.044) & \textbf{0.664} (0.064) & \textbf{0.625} (0.067) & \textbf{0.516} (0.063) \\
    Copula (20) & 0.752 (0.056) & 0.593 (0.068) & 0.562 (0.066) & 0.495 (0.065) \\
    Copula (30) & 0.676 (0.059) & 0.557 (0.062) & 0.518 (0.063) & 0.457 (0.070) \\
    \hline
    \hline
    \textbf{Model}  & $\textbf{n = 1000, p = 50}$ & $\textbf{n = 2000, p = 50}$ & $\textbf{n = 5000, p = 50}$ & $n = \textbf{10000, p = 50}$  \\
    \hline
    Subbotin & \textbf{0.612} (0.080) & \textbf{0.647} (0.074) & 0.657 (0.068) & 0.712 (0.065) \\
    \hline
    Gaussian NS & 0.480 (0.071) & 0.496 (0.062) & 0.521 (0.064) & 0.563 (0.064) \\
    Gaussian Glasso & 0.495 (0.068) & 0.506 (0.062) & 0.517 (0.067) & 0.541 (0.067) \\
    \hline
    Quantile (0.5) & $<$ 0.001 ($<$ 0.001) & $<$ 0.001 ($<$ 0.001) & $<$ 0.001 ($<$ 0.001) & $<$ 0.001 ($<$ 0.001) \\
    Quantile (0.9) & 0.267 (0.061) & 0.285 (0.061) & 0.308 (0.060) & 0.341 (0.074) \\
    Quantile (0.99) & 0.369 (0.064) & 0.373 (0.062) & 0.397 (0.064) & 0.434 (0.069) \\
    \hline
    Copula (10) & 0.608 (0.068) & 0.642 (0.061) & \textbf{0.664} (0.064) & \textbf{0.733} (0.063) \\
    Copula (20) & 0.605 (0.063) & 0.616 (0.065) & 0.643 (0.068) & 0.642 (0.064) \\
    Copula (30) & 0.545 (0.070) & 0.576 (0.072) & 0.597 (0.062) & 0.611 (0.060) \\
    \hline
\end{tabular}
\end{footnotesize}
\end{table}

\subsection{Block Maxima Extreme Value Distribution}

Table \ref{tab:bm} shows the results from the block maxima extreme value simulations with data-driven model selection via stability selection. For each simulation set, we show the average and standard deviation of the F1 score for each model with respect to estimating the true edge set of the underlying graph of the data across 50 replications. We resample the original data with the same number of observations as the original data created using block bootstrapping with random block lengths, and we use a stability threshold of 0.9. Here, we use stability selection to select both $\nu$ and $\lambda$ of the Subbotin graphical model; $\nu = 4$ was by far the most common value of $\nu$ selected for the Subbotin graphical model. In general, the block maxima copula graphical model still generally performs best in the simulations. However, the difference between the F1 scores of the models are much smaller the oracle sparsity tuning results. This is likely because of the resampling of the observations used for stability selection; while the block bootstrapping procedure preserves some of the time-based structure in the data, some of the blocks will be divided. Thus, some of the block maxima in the resampled data may not be extreme values, or there may be multiple extreme values in a single block.  The quantile and Gaussian graphical models both show a similar small decrease in F1 scores compared to oracle sparsity tuning.

\subsection{Peaks-Over-Threshold Extreme Value Distribution}

Table \ref{tab:pot} shows the results from the peaks-over-threshold extreme value simulations with data-driven model selection via stability selection. For each simulation set, we show the average and standard deviation of the F1 score for each model with respect to estimating the true edge set of the underlying graph of the data across 50 replications. We resample the original data with the same number of observations as the original data created using block bootstrapping with random block lengths, and we use a stability threshold of 0.9. We use stability selection to select both $\nu$ and $\lambda$ for the Subbotin graphical model. All models perform worse compared to the oracle sparsity tuning results, which is a reflection of the possibility that a portion of the rarely occurring extreme values may not be part of the resampled data created the bootstrap procedure. Thus, there may not be observed correlated extreme values between two features with an edge in the underlying graph. The relative performance of each of the models is similar to what is seen in the oracle sparsity tuning results. The Subbotin graphical model has the highest average F1 scores across the simulation scenarios, with a value of $\nu$ of 4 or 6 always chosen by stability selection, while the quantile graphical model at a high quantile and the Gaussian graphical model also perform decently. The block maxima copula graphical model does not do well in this case, as was the case in the oracle sparsity tuning.

\begin{table}[t]
\caption{Average F1 scores (std. dev.) for the peaks-over-threshold simulations with data-driven edge selection, averaged over 50 replicates. Best performing methods are boldfaced.}
\label{tab:pot}
\centering
\begin{footnotesize}
\begin{tabular}{|l|r|r|r|r|}
    \hline
    \textbf{Model}  & $\textbf{n = 10000, p = 25}$ & $\textbf{n = 10000, p = 50}$ & $\textbf{n = 10000, p = 75}$ & $\textbf{n = 10000, p = 100}$ \\
    \hline
    Subbotin & \textbf{0.706} (0.032) & \textbf{0.674} (0.040) & \textbf{0.635} (0.049) & \textbf{0.592} (0.057) \\
    \hline
    Gaussian NS & 0.597 (0.030) & 0.503 (0.037) & 0.487 (0.044) & 0.460 (0.049) \\
    Gaussian Glasso & 0.602 (0.031) & 0.504 (0.035) & 0.482 (0.048) & 0.455 (0.045) \\
    \hline
    Quantile (0.5) & $<$ 0.001 ($<$ 0.001) & $<$ 0.001 ($<$ 0.001) & $<$ 0.001 ($<$ 0.001) & $<$ 0.001 ($<$ 0.001) \\
    Quantile (0.9) & 0.562 (0.036) & 0.515 (0.042) & 0.461 (0.051) & 0.357 (0.063) \\
    Quantile (0.99) & 0.647 (0.035) & 0.619 (0.048) & 0.548 (0.057) & 0.523 (0.060) \\
    \hline
    Copula (10) & 0.178 (0.021) & 0.063 (0.016) & 0.065 (0.013) & 0.042 (0.010) \\
    Copula (20) & 0.346 (0.035) & 0.221 (0.014) & 0.157 (0.012) & 0.156 (0.011) \\
    Copula (30) & 0.318 (0.013) & 0.124 (0.012) & 0.115 (0.014) & 0.098 (0.015) \\
    \hline
    \hline
    \textbf{Model}  & $\textbf{n = 5000, p = 50}$ & $\textbf{n = 7500, p = 50}$ & $\textbf{n = 10000, p = 50}$ & $\textbf{n = 12500, p = 50}$  \\
    \hline
    Subbotin & \textbf{0.571} (0.067) & \textbf{0.629} (0.058) & \textbf{0.674} (0.049) & \textbf{0.706} (0.044) \\
    \hline
    Gaussian NS & 0.391 (0.048) & 0.480 (0.037) & 0.503 (0.037) & 0.616 (0.037) \\
    Gaussian Glasso & 0.397 (0.043) & 0.471 (0.041) & 0.504 (0.035) & 0.615 (0.036) \\
    \hline
    Quantile (0.5) & $<$ 0.001 ($<$ 0.001) & $<$ 0.001 ($<$ 0.001) & $<$ 0.001 ($<$ 0.001) & $<$ 0.001 ($<$ 0.001) \\
    Quantile (0.9) & 0.367 (0.057) & 0.448 (0.044) & 0.515 (0.042) & 0.617 (0.037) \\
    Quantile (0.99) & 0.488 (0.068) & 0.534 (0.057) & 0.619 (0.048) & 0.89 (0.033) \\
    \hline
    Copula (10) & 0.082 (0.017) & 0.084 (0.020) & 0.063 (0.016) & 0.116 (0.023) \\
    Copula (20) & 0.146 (0.015) & 0.205 (0.022) & 0.221 (0.014) & 0.256 (0.025) \\
    Copula (30) & 0.053 (0.014) & 0.115 (0.020) & 0.124 (0.012) & 0.147 (0.022) \\
    \hline
\end{tabular}
\end{footnotesize}
\end{table}

\newpage

\section{Proof of Theorem 1}

Our derivation of the functional form of the joint distribution follows from \cite{jb1}, which showed that a valid multivariate joint distribution can be derived from a known conditional distribution via the equivalency $$Q(\xx) = \prod_{i = 1}^p \frac{P(x_{i}|x_1, x_2, \hdots, x_{i-1}, y_{i+1}, \hdots y_p)}{P(y_{i}|x_1, x_2, \hdots, x_{i-1}, y_{i+1}, \hdots y_p )}$$ (Note that since the distribution is supported on the entirety of $\mathbb{R}^p$, the necessary assumption of a non-zero probability for all denominators is fulfilled.) We derive the exponential functional form of the joint distribution in Theorem 1 by defining $$Q(\xx) := \log(P(\xx) / P(\boldsymbol{0})),$$ which can be shown to be equivalent to $$Q(\xx) = \log \left(\prod_{i = 1}^p \frac{P(x_{i}|x_1, x_2, \hdots, x_{i-1}, y_{i+1} = 0, \hdots y_p = 0)}{P(y_{i} = 0|x_1, x_2, \hdots, x_{i-1}, y_{i+1} = 0, \hdots y_p = 0)}\right).$$ We then replace $Q(\xx)$ with the Hammersley-Clifford definition of the most general form of a graphical model joint distribution to get: $$ \sum_{1 \leq i \leq p} x_i G_i(x_i) + \sum_{1 \leq i \leq j \leq p} x_i x_j G_{ij}(x_i, x_j) + \hdots  = \log \left(\prod_{i = 1}^p \frac{P(x_{i}|x_1, x_2, \hdots, x_{i-1}, y_{i+1} = 0, \hdots y_p = 0)}{P(y_{i} = 0|x_1, x_2, \hdots, x_{i-1}, y_{i+1} = 0, \hdots y_p = 0)}\right).$$ For our particular case, there will be interaction terms in the Hammersley-Clifford form of the distribution up to the $\nu$-th order. We can then find equivalencies between individual terms on the left and right hand sides of the above equation. Without loss of generality, we first operate on $\xx_p.$ Define: $$\xx_{i:0} := (x_1, \hdots, x_{i-1}, 0, x_{i + 1}, \hdots, x_p).$$ By definition, we have $$\frac{\exp(Q(\xx))}{\exp(Q(\xx_{p:0}))} = \exp(Q(\xx) - Q(\xx_{p:0})) = x_p\left(G_p(x_p) + \sum_{1 \leq i \leq p-1} x_i G_{ip}(x_i, x_p)\right)$$ and $$\frac{\exp(Q(\xx))}{\exp(Q(\xx_{p:0}))} = \frac{P(\xx)}{P(\xx_{p:0})} = \frac{P(x_{p}|x_1, x_2, \hdots, x_{p-1})}{P(x_{p} = 0|x_1, x_2, \hdots, x_{p-1})}.$$ The two above equations allows us to isolate only those terms which are functions of $x_p$. The latter equation also gives us the first term of the product $\prod_{i = 1}^p \frac{P(x_{i}|x_1, x_2, \hdots, x_{i-1}, y_{i+1} = 0, \hdots y_p = 0)}{P(y_{i} = 0|x_1, x_2, \hdots, x_{i-1}, y_{i+1} = 0, \hdots y_p = 0)}.$ Specifically we get: 
\begin{eqnarray*}
    x_p\left(G_p(x_p) + \sum_{1 \leq i \leq p-1} x_i G_{ip}(x_i, x_p)  + \hdots\right) & = & \frac{P(x_{p}|x_1, x_2, \hdots, x_{p-1})}{P(x_{p} = 0|x_1, x_2, \hdots, x_{p-1})}\\
    & = & \exp\left(-(x_p - \sum_{i \leq p-1} \theta_{ip}x_i)^{\nu}  + (\sum_{i \leq p-1} \theta_{ip}x_i)^{\nu}\right) \\
\end{eqnarray*} Using arithmetic to expand the right hand side of the above equation, we can find each individual $G$ function: $$G_p(x_p) = - x_p^{\nu - 1}, $$ $$G_{ip}(x_i, x_p) =  - \theta_{ip} \left((x_p - \theta_{ip} x_i)^{\nu - 2} - (\theta_{ip} x_i)^{\nu - 1} \right),$$ $$ G_{ijp}(x_i, x_j, x_p) = - \theta_{ip} \theta_{jp} \left((x_p - \theta_{ip} x_i - \theta_{jp} x_j)^{\nu - 3}  - (\theta_{ip} x_i - \theta_{jp} x_j)^{\nu - 2} \right),$$  and so on, up to the $\nu$-th order $G$ function. One particular thing to note is that, if $\theta_{ip} = 0$, then all $G$ functions involving $x_i$ and $x_p$ will be equal to 0, meaning that this is a valid form for the joint probability distribution by the Hammersley-Clifford theorem. We can then operate recursively to get the $G$ functions for all $r \in 1, 2, \hdots, p$. For example, for $x_{p-1}$, we get: \begin{eqnarray*}
    x_{p-1}\left(G_{p-1}(x_{p-1}) + \sum_{1 \leq i \leq p-2} x_i G_{i(p-1)}(x_i, x_p)  + \hdots\right) & = & \frac{P(x_{p-1}|x_1, x_2, \hdots, x_{p-2}, x_p = 0)}{P(x_{p-1} = 0|x_1, x_2, \hdots, x_{p-2}, x_p = 0)}\\
    & = & \exp\left(-(x_p - \sum_{i \leq p-2} \theta_{ip}x_i)^{\nu}  + (\sum_{i 
\leq p-2} \theta_{ip}x_i)^{\nu}\right) \\
\end{eqnarray*} Putting all terms together for all variables gives us following functional form for the joint distribution:
\begin{eqnarray*}
    Q(\xx) & = & \log \left(\prod_{i = 1}^p \frac{P(x_{i}|x_1, x_2, \hdots, x_{i-1}, y_{i+1} = 0, \hdots y_p = 0)}{P(y_{i} = 0|x_1, x_2, \hdots, x_{i-1}, y_{i+1} = 0, \hdots y_p = 0)}\right) \\
    & = & \sum_{i = 1}^p \left( - (\theta_{ii} x_i - \sum_{j < i} \theta_{ij}x_j)^{\nu} + (\sum_{j < i} \theta_{ij}x_j)^{\nu} \right)\\
    f(\xx) & \propto& \exp(Q(\xx)) \\
    & \propto& \exp\left( \sum_{i = 1}^p \left( - (\theta_{ii}x_i - \sum_{j < i} \theta_{ij}x_j)^{\nu} + (\sum_{j < i} \theta_{ij}x_j)^{\nu} \right)\right).
\end{eqnarray*} Adding the log normalizing constant to the distribution gives us:
\begin{eqnarray*}
    f(\xx) & = & \exp \left(\sum_{i = 1}^p \left( -(\theta_{ii}x_i - \sum_{j < i} \theta_{ij}x_j)^{\nu} + (\sum_{j < i} \theta_{ij}x_j)^{\nu} \right) - A(\boldsymbol{\Theta}) \right). 
\end{eqnarray*}

\hfill $\square$

\section{Proof of Theorem 2}

We first discuss the relationship between $\boldsymbol{\Theta}$ and the covariance of the distribution. Let $\boldsymbol{\Sigma}$ be the covariance matrix of the Subbotin graphical model distribution. By the definition of the node-wise conditional distributions, $\boldsymbol{\Theta}$ contains the partial correlations between all pairs of variables of the multivariate distribution. We can then combine the results from for the relationship between the partial correlation and precision matrices \cite{sl1} with the results from for the moments of univariate Subbotin distribution \cite{sn1}, in order to find the relationship between $\boldsymbol{\Sigma}$ and $\boldsymbol{\Theta}$: $$\boldsymbol{\Sigma}^{-1}_{ij} = \begin{cases} 
    \left(\frac{\Gamma(1/\nu)}{\Gamma(3/\nu)}\right) \theta_{ii} & i = j \\
    - \left(\frac{\Gamma(1/\nu)}{\Gamma(3/\nu)}\right) \theta_{ij} & i \neq j \\
  \end{cases}.$$ Therefore, because the empirical $\boldsymbol{\Sigma}$ must be positive definite by definition for any data set, the empirical $\boldsymbol{\Theta}$ must be positive definite for any data set as well.

We now show that the condition in Theorem 2 is a necessary and sufficient condition for normalizability. Our proof requires the following lemma:

\begin{lem}
\label{lem:ass} 
Define $$Q_{i,\nu}(x_i) = \left( -(\theta_{ii}x_i - \sum_{j < i} \theta_{ij}x_j)^{\nu} + (\sum_{j < i} \theta_{ij}x_j)^{\nu} \right)$$ where $\nu$ is an even integer greater than 2. Then, for all $i \in \{1, 2, \hdots, p\}$, 

\begin{enumerate}
    \item $sign(Q_{i,2}(x_i)) = sign(Q_{i,\nu}(x_i)).$
    \item $\lim_{x_i \rightarrow \pm \infty}  \frac{Q_{i,2}(x_i)}{Q_{i,\nu}(x_i)} = 0.$
\end{enumerate}
\end{lem} 

\noindent To prove the first part of Lemma \ref{lem:ass}, we exhaustively examine all possibilities for $sign(Q_{i,2}(x_i))$. We first note that $$Q_{i,\nu}(x_i) = \left( -(\theta_{ii}x_i - \sum_{j < i} \theta_{ij}x_j)^{\nu} + (\sum_{j < i} \theta_{ij}x_j)^{\nu} \right)$$ $$ = \left( -((\theta_{ii}x_i - \sum_{j < i} \theta_{ij}x_j)^{2})^{\nu/2} + ((\sum_{j < i} \theta_{ij}x_j)^{2})^{\nu/2} \right).$$  If $sign(Q_{i,2}(x_i)) = +1$, then $$(\theta_{ii}x_i - \sum_{j < i} \theta_{ij}x_j)^{2} < (\sum_{j < i} \theta_{ij}x_j)^{2}. $$ It then follows that $$((\theta_{ii}x_i - \sum_{j < i} \theta_{ij}x_j)^{2})^{\nu/2} < ((\sum_{j < i} \theta_{ij}x_j)^{2})^{\nu/2}$$ since the transformation operator in this case is monotonically increasing. Thus, in this case, we have that $sign(Q_{i,\nu}(x_i)) = +1$ as well.
Similarly, if $sign(Q_{i,2}(x_i)) = -1$, then we have $$(\theta_{ii}x_i - \sum_{j < i} \theta_{ij}x_j)^{2} > (\sum_{j < i} \theta_{ij}x_j)^{2}.$$ Using the same argument as above, this implies that $$((\theta_{ii}x_i - \sum_{j < i} \theta_{ij}x_j)^{2})^{\nu/2} > ((\sum_{j < i} \theta_{ij}x_j)^{2})^{\nu/2}.$$ Therefore, in this scenario, $sign(Q_{i,\nu}(x_i)) = -1$ as well. The second part of Lemma \ref{lem:ass} can be shown simply by direct comparison:

\begin{eqnarray*}
 && \lim_{x_i \rightarrow \pm \infty}  \frac{Q_{i,2}(x_i)}{Q_{i,\nu}(x_i)} \\
 & = & \lim_{x_i \rightarrow \pm \infty}  \frac{1}{-(\theta_{ii}x_i - \sum_{j < i} \theta_{ij}x_j)^{\nu - 2}} + o(1) \\
 & = & 0
\end{eqnarray*}

We now use the statement of Lemma \ref{lem:ass} to show necessity and sufficiency for the condition of Theorem 2. To do this, we utilize the known normalizability condition of the Subbotin graphical model for $\nu = 2$, i.e. when the Subbotin graphical model is equivalent to a multivariate Gaussian distribution. Define $$Q_{\nu}(\xx) = \sum_{i = 1}^p \left( -(\theta_{ii}x_i - \sum_{j < i} \theta_{ij}x_j)^{\nu} + (\sum_{j < i} \theta_{ij}x_j)^{\nu} \right).$$ If $\boldsymbol{\Theta}$ is positive definite, then we know from previous results on the multivariate Gaussian that $\forall \, i \in \{1, 2, \hdots, p\}$, $$\lim_{x_i \rightarrow \pm \infty} \exp(Q_{i, 2}(\xx)) = 0 $$ and therefore $$\int_{-\infty}^{\infty} \int_{-\infty}^{\infty} \int_{-\infty}^{\infty} \hdots  \exp(Q_{2}(\xx)) d x_1 d x_2 d x_3 \hdots$$ is finite. From Lemma \ref{lem:ass}, it can be shown that for any even integer $\nu$ greater than 2:  $$\forall i \in \{1, 2, \hdots, p\}: \lim_{x_i \rightarrow \infty}  \frac{\exp( Q_{i, \nu}(\xx))}{\exp( Q_{i, 2}(\xx))} = 0.$$ Thus, we have that $\forall \, i \in \{1, 2, \hdots, p\}$, $$\lim_{x_i \rightarrow \pm \infty} \exp(Q_{i, \nu}(\xx)) = 0$$ and therefore $$\int_{-\infty}^{\infty} \int_{-\infty}^{\infty} \int_{-\infty}^{\infty} \hdots  \exp(Q_{\nu}(\xx)) d x_1 d x_2 d x_3 \hdots$$ will be finite as well. Thus, the condition of Theorem 2 is a sufficient condition for normalizability. On the other hand, if $\boldsymbol{\Theta}$ is not positive definite, then $\exists \, i \in \{1, 2, \hdots, p\}$, $$\lim_{x_i \rightarrow \pm \infty} \exp(Q_{i, 2}(\xx)) = \infty. $$ Therefore, $$\int_{-\infty}^{\infty} \int_{-\infty}^{\infty} \int_{-\infty}^{\infty} \hdots  \exp(Q_{2}(\xx)) d x_1 d x_2 d x_3 \hdots$$ will be divergent. From Lemma \ref{lem:ass}, we have that for any even integer $\nu$ greater than 2:  $$\forall i \in \{1, 2, \hdots, p\}: \lim_{x_i \rightarrow \infty}  \frac{\exp(Q_{i, 2}(\xx))}{\exp(Q_{i, \nu}(\xx))} = \infty.$$ Thus, by the limit comparison test,  $\exists \, i \in \{1, 2, \hdots, p\}$ such that $$\lim_{x_i \rightarrow \pm \infty} Q_{i, \nu}(x_i) < 0$$ and $$\int_{-\infty}^{\infty} \int_{-\infty}^{\infty} \int_{-\infty}^{\infty} \hdots  \exp(-Q_{\nu}(\xx)) d x_1 d x_2 d x_3 \hdots$$ will be divergent as well, meaning that $A(\boldsymbol{\Theta})$ will not exist. Thus, the condition of Theorem 2 is a necessary condition for normalizability. 

\hfill $\square$

\section{Proof of Theorem 3}

To prove this statement, we apply previous theoretical results for edge selection consistency for the Extreme Lasso regression which we utilize as the node-wise conditional regression problem for the Subbotin graphical model. We first provide the statement for the finite sample convergence rate and variable selection consistency of the Extreme Lasso problem; this corresponds to Theorem 3.4 in \citep{ac2}; the full proof can be found in the aforementioned paper.

\begin{lem}[\textbf{Neighborhood Selection Consistency}]  \label{elconsis} Consider the Extreme Lasso program for the node-wise conditional regression problem.  Assume that conditions 1 and 2 from section 3 hold. Consider the family of regularization parameters $\lambda =  \frac{4 \kappa_{\text{IC}}}{\tau} \gamma \sqrt{ \frac{\log p}{n}} \bigg[    2 \sqrt{ \frac{2}{\gamma}}   +   \sqrt{ \frac{\log p}{n}}   \bigg] $,  where $\kappa_{\text{IC}}$ is the  compatibility constant defined in \citet{lee}. Then the following properties holds with probability greater than   $1 - c_1 \exp(- c_2^* \log p)$: 

(i) The Lasso has a unique solution with support contained
within $S_i$ i.e., $\hat{S}_i \subset S_i$.

(ii) If condition 3 holds, the lasso estimator is also sign consistent: $\text{sign} (\hat \theta_{ij}) = \text{sign} (\theta_{ij})$.
\end{lem}

\noindent The full proof can be found in the aforementioned paper. From Lemma \ref{elconsis}, we can use a simple union bound on all node neighborhoods to derive the probability of model selection consistency for the full graph. If variable selection consistency holds for a single neighborhood with probability $$1 - c_1 \exp(- c_2^* \log p),$$ then it does not hold with probability $$c_1 \exp(- c_2^* \log p).$$ Thus, the probability of any neighborhood not holding will be bounded above by $$p c_1 \exp(- c_2^* \log p).$$ Therefore, the probability of achieving consistent model selection for the entire graph will be bounded below by $$1 - c_1 \exp(- c_2^* \log p + \log p).$$ We can then define $c_2 = c_2^* - 1$ to get $$1 - c_1 \exp(- c_2 \log p).$$

\hfill$\square$

\label{lastpage}

\end{document}